\documentclass{article}
\usepackage{microtype}
\usepackage{graphicx}
\usepackage{subfigure}
\usepackage{booktabs} 

\usepackage{csquotes}
\usepackage{algorithm2e}
\RestyleAlgo{ruled}
\SetKwComment{Comment}{/* }{ */}

\usepackage{hyperref}

\usepackage[accepted]{iclr2024-latex/icml2024}

\icmltitlerunning{Submission and Formatting Instructions for ICML 2024}

\usepackage{amsmath}
\usepackage{amssymb}
\usepackage{mathtools}
\usepackage{amsthm}
\usepackage[version=4]{mhchem}

\usepackage[capitalize,noabbrev]{cleveref}

\theoremstyle{plain}

\theoremstyle{definition}

\theoremstyle{remark}

\usepackage{booktabs}
\usepackage{siunitx}

\usepackage{enumitem}

\usepackage[textsize=tiny]{todonotes}

\NewDocumentCommand{\carl}{mO{}}{\textcolor{green}{\textsuperscript{\textit{Carl}}\textsf{\textbf{\small[#1]}}}}

\newcommand{\ourSystem}{\textsc{ChemReasoner}}
\newcommand{\argtopk}[1]{\mathop{\mathrm{arg\,top}\,#1}}
\icmltitlerunning{$\ourSystem$: Heuristic Search over an LLM’s Knowledge Space using Quantum-Chemical Feedback}
\begin{document}

\twocolumn[

\icmltitle{$\ourSystem$: Heuristic Search over a Large Language Model’s\\ Knowledge Space using Quantum-Chemical Feedback}



\icmlsetsymbol{equal}{*}

\begin{icmlauthorlist}

\icmlauthor{Henry W. Sprueill}{pnnl}
\icmlauthor{Carl Edwards}{uiuc}
\icmlauthor{Khushbu Agarwal}{pnnl}
\icmlauthor{Mariefel V. Olarte}{pnnl}
\icmlauthor{Udishnu Sanyal}{pnnl}
\icmlauthor{Conrad Johnston}{msft}
\icmlauthor{Hongbin Liu}{msft}
\icmlauthor{Heng Ji}{uiuc}
\icmlauthor{Sutanay Choudhury}{pnnl}
\end{icmlauthorlist}

\icmlaffiliation{pnnl}{Pacific Northwest National Laboratory}
\icmlaffiliation{uiuc}{University of Illinois Urbana-Champaign,}
\icmlaffiliation{msft}{Azure Quantum, Microsoft}

\icmlcorrespondingauthor{Sutanay Choudhury}{sutanay.choudhury@pnnl.gov}

\icmlkeywords{Machine Learning, ICML}

\vskip 0.3in
]



\printAffiliationsAndNotice{}  

\begin{abstract}

The discovery of new catalysts is essential for the design of new and more efficient chemical processes in order to transition to a sustainable future. We introduce an AI-guided computational screening framework unifying linguistic reasoning with quantum-chemistry based feedback from 3D atomistic representations. Our approach formulates catalyst discovery as an uncertain environment where an agent actively searches for highly effective catalysts via the iterative combination of large language model (LLM)-derived hypotheses and atomistic graph neural network (GNN)-derived feedback. Identified catalysts in intermediate search steps undergo structural evaluation based on spatial orientation, reaction pathways, and stability. Scoring functions based on adsorption energies and reaction energy barriers steer the exploration in the LLM's knowledge space toward energetically favorable, high-efficiency catalysts. We introduce planning methods that automatically guide the exploration without human input, providing competitive performance against expert-enumerated chemical descriptor-based implementations. By integrating language-guided reasoning with computational chemistry feedback, our work pioneers AI-accelerated, trustworthy catalyst discovery.\footnote{The code and datasets discussed in this paper are available at \url{https://github.com/pnnl/chemreasoner}.}

\end{abstract}

\section{Introduction}

The discovery of new catalysts requires one to identify the optimal combination of chemical descriptors (or properties) and use these descriptors to propose catalysts. However, such descriptors are only empirically understood, presenting a challenge for computational studies of catalysis.
Typically, chemists actively reason to mentally search through reactants, catalysts, and operating conditions that enable more energy-efficient chemical conversions. 
However, as discussed by \citet{norskov2011density}, linking microscopic surface properties to macroscopic catalytic performance via chemical descriptors remains a barrier to descriptor-based catalyst search.

Large Language Models (LLMs) \cite{wei2022chain, ouyang2022training, taylor2022galactica, BiomedicalLM2023, openai2023gpt4} offer a new opportunity to realize such a data-driven autonomous search to accelerate scientific discovery. Our work aims to enhance natural language reasoning capabilities with quantum-chemical feedback to discover optimal catalysts for target reactions. 

\begin{figure*}[h]
  \centering
  \includegraphics[width=0.9\textwidth]{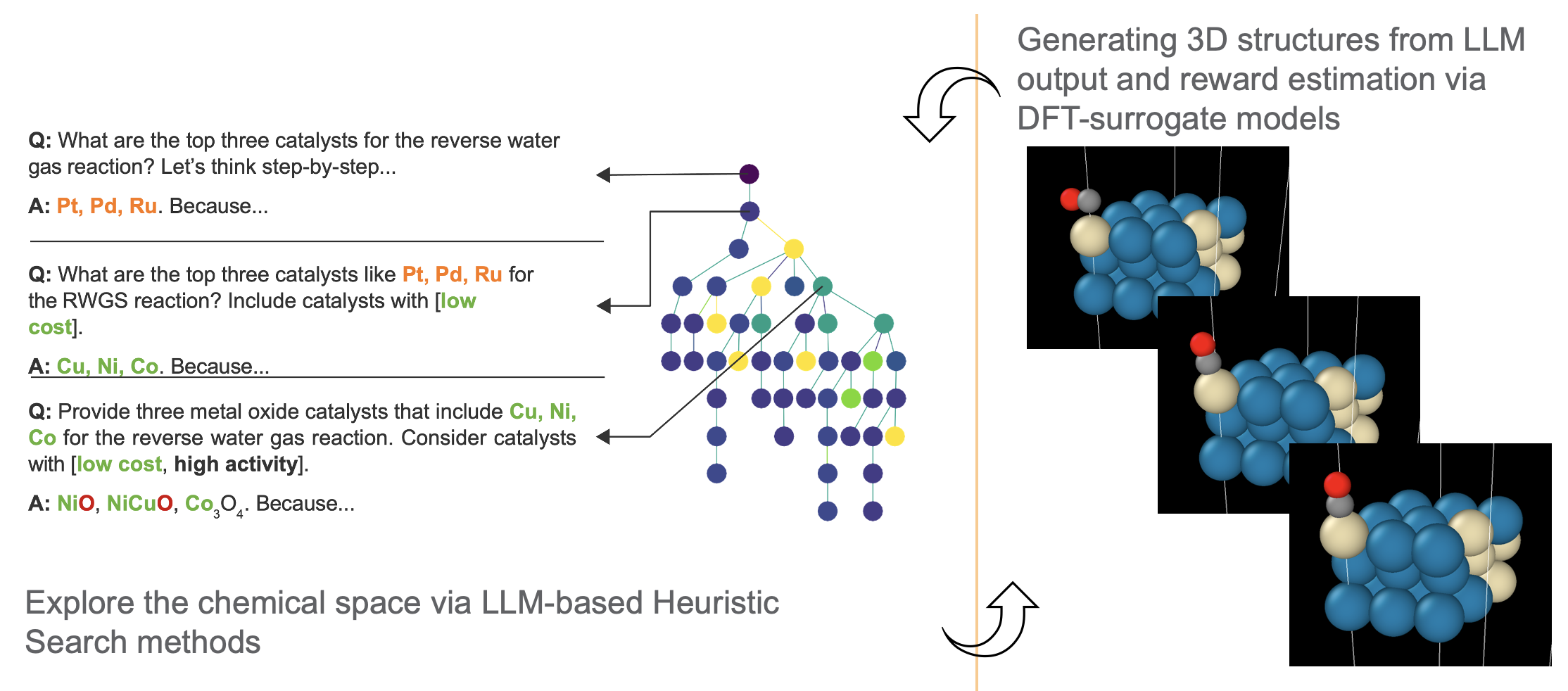}
\caption{ChemReasoner successively ``thinks'' in terms of different constraints and factors, which are based on scientific principles and narrow down the set of possible candidates.  It accomplishes that by prompting a language model with different combinations of chemical descriptors, yielding a tree-structured space of queries and potential candidates, and returns the optimal answer via efficient exploration of the search-space. ChemReasoner uses automated planning, based on previous reasoning, to initiate the exploration and guides it via a reward obtained from the exploration process to prune unpromising actions. We currently
use “adsorption energy”, a key measure of reactivity as the reward function.}
\label{fig:challenge_and_approach}
\end{figure*}

\textbf{The Challenge} However, reasoning about complex catalytic processes requires modelling across multiple modalities, extending beyond the capabilities of existing language models. This includes merging scientific concepts from literature and property prediction with 3D atomistic configurations. Determining the best catalyst is a multi-step process, requiring reasoning about multiple macroscopic properties. The first step involves identifying an optimal set of chemical descriptors (e.g. ``resistance to poisoning", ``porosity") which are relevant to the reaction in question. Formally, given a set of these important descriptors $\mathbf{P},|\mathbf{P}|=n$ , we want to identify the optimal subset $\mathbf{R\subset \mathbf{P}},|\mathbf{R}|=r$ to consider when suggesting new catalysts. This yields $P^{n}_{r}=\frac{n!}{(n-r)!}$ possible permutations to reason over. 
The amount of reasoning required scales combinatorially with the number of available properties, necessitating an autonomous reasoner: an LLM. Using its knowledge of scientific concepts, the LLM both proposes important properties and proposes the best possible catalysts (from all possible catalysts) that have these properties. 
Pruning this large space of candidate catalysts requires reasoning about the complex microscopic interactions that occur between atomistic structures in 3D space based on macroscopic properties (Figure \ref{fig:search_tree}). Further, while simple reactions can be assessed via adsorption energies of 3D chemical structures, complex reactions demand consideration of multi-step reaction pathways and selectivity \cite{unsleber2023high}.

\textbf{Technical Approach} To solve this challenge, we propose a framework that combines LLM-driven heuristic search for catalyst discovery with structure-based scoring from atomistic graph neural networks (GNNs) trained from quantum chemistry simulations for guidance (Fig. \ref{fig:challenge_and_approach}). This framework formulates catalyst discovery as an uncertain environment where an agent (the LLM) pursues energetically favorable catalysts based on computational chemistry feedback. In each search step, the agent plans its actions\cite{huang2022inner, hao2023reasoning} by 1) automatically identifying the optimal set of properties to consider, 2) generating new search prompts based on the identified properties, and 3) executing the prompts using sophisticated instruction following \cite{ouyang2022training}. Catalyst candidates identified in each step of the search are transformed into 3D atomistic representations of the catalyst-adsorbate structure \cite{zitnick2020introduction}. These representations enable evaluation via structural evaluation -- including spatial orientation, energy barriers over reaction pathways, and stability -- yielding a reward for 
catalyst suitability. 
This reward drives the LLM towards catalysts which enable reactions with minimal external energy, a crucial step for developing environmentally friendly industrial processes. 

\noindent In this work, we make the following key contributions:
\begin{enumerate}[itemsep=0pt,parsep=1pt,topsep=0pt,partopsep=0pt]
    \item We introduce $\ourSystem$: a novel hypothesis generation and testing framework unifying heuristic search over an LLM's knowledge-space with quantum chemistry-guided feedback. This enables natural language-based reasoning for catalyst discovery with stronger domain guarantees obtained from computational chemistry methods. 
    \item We demonstrate the decisive impact of planning methods in automatically navigating chemical search spaces over a SOTA LLM-based implementation. Our purely LLM-planned approach  with zero human input ($\ourSystem$\text{-Planner}) surpasses search guided by expert-selected chemical descriptors ($\ourSystem$\text{-Expert}) for two out of three categories in our evaluation benchmark.
    \item Third, and uniquely, we establish the domain-grounding of language models via quantum chemical property feedback. We go beyond screening catalysts on adsorption energies alone and propose a methodology to reason in terms of reaction pathways and energy barriers.
\end{enumerate}

Our work pioneers an AI-guided approach to computational catalyst screening and discovery. To facilitate community adoption and advancement of this extremely interdisciplinary and compute-intensive research, all of our datasets and code are freely available on github. This includes query benchmarks from catalysis experts, multi-modal provenance trails, over 700,000+ atomistic trajectories, and additional validation of catalyst candidates from density functional theory calculations.

\section{Background and Related Work}

\subsection{Catalysis}
Catalysts accelerate chemical reactions by lowering reaction barriers, without being consumed in the process. Heterogeneous catalysis, wherein the catalyst is in a different phase than the reactants and products, is widely used in industrial chemical processes \cite{dumesic2008principles}. Developing novel heterogeneous catalysts with high activity and selectivity is essential to design energy efficient chemical process that paves a way towards sustainability \cite{zitnick2020introduction, hu2021heterogeneous, mukhtar2022current}.

In heterogeneous catalysis, gases or liquids interact with a solid catalyst surface to enable a reaction \cite{greeley2002electronic}. This overall process consists of three elementary steps: 1) Adsorption - reactant molecules bind to the catalyst surface; 2) Surface reaction - adsorbed molecules react with each other to generate product(s); 3) Desorption - product(s) molecules desorb from the catalyst surface.

Adsorption energy of a reactant, reflecting its binding strength on a specific catalyst surface, is often identified as one of the  key criterion for the activity of that catalyst. Fundamentally, adsorption energy of a reactant can be tuned by changing  microscopic properties such as surface structure (crystal facet)  or electronic configuration. Within catalysis, an overarching goal is to link these microscopic descriptors to macroscopic catalytic performance metrics to allow computation descriptor-based catalyst search \cite{norskov2011density}. 
However,  modeling complex catalyst surface reactions requires going beyond just adsorption descriptors \cite{xi2002situ}. Interactions between reaction intermediates and competition between multiple possible reaction pathways must be considered as well \cite{kattel2017tuning, schwaller2019molecular, chen2022generalized, unsleber2023high}. With so many interactions in mind, we pursue a goal of developing models to reason compositionally about descriptors, structures, and pathways to generate high-quality hypotheses for potential catalysts, applied to the production of sustainable fuels.
\begin{figure*}[!h]
  \centering
  \includegraphics[width=0.7\textwidth]{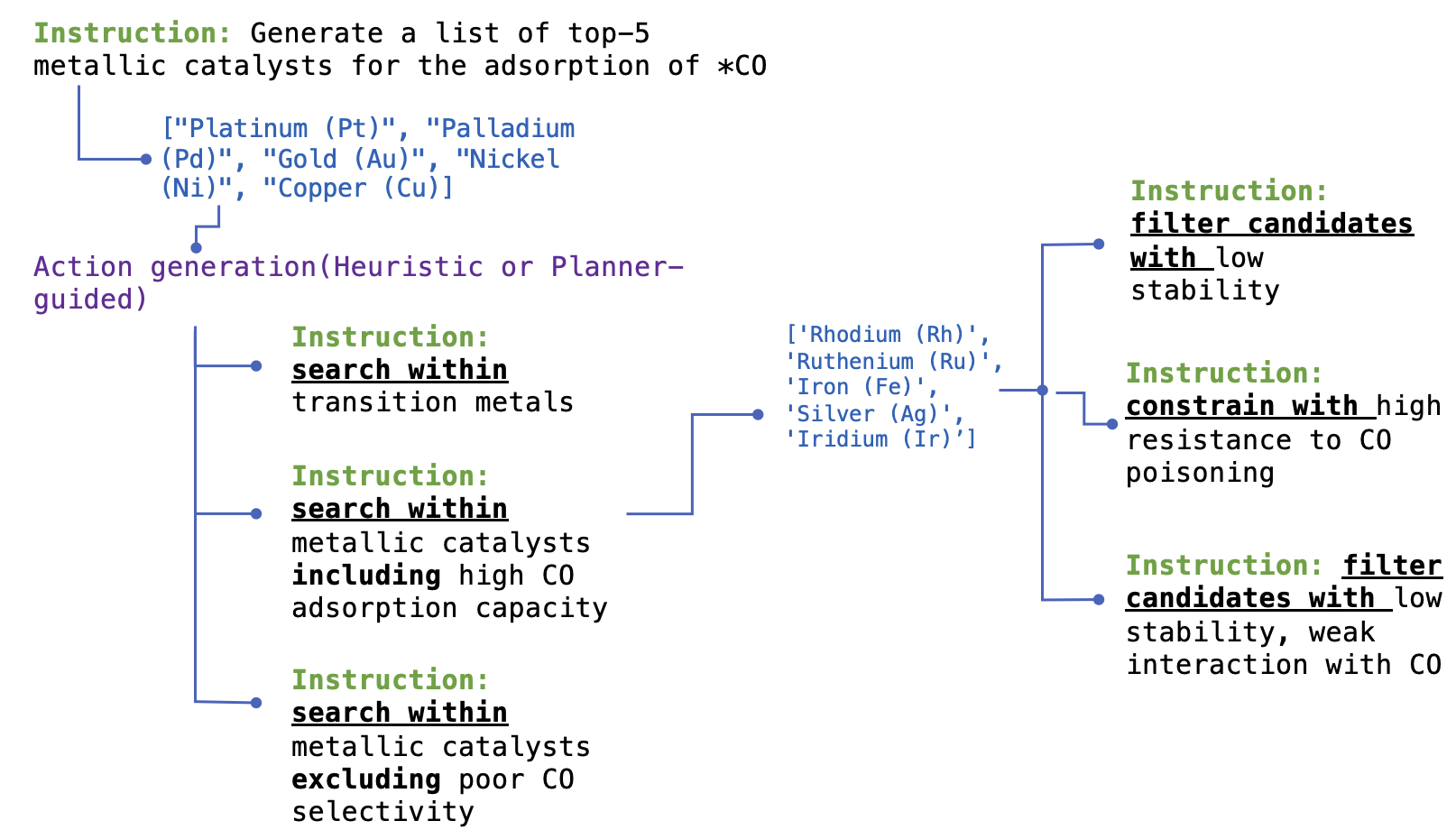}
\caption{Illustration of {\ourSystem} search process (best viewed in color): The initial question generates base candidates, which are iteratively refined by adding an optimal set of constraints to the query and producing a new set of actions (or prompts) to explore the LLMs internal knowledge space. The optimal action set is chosen by 1) sampling from expert specified action space, or 2) automated generated by a planner component as illustrated in Figure \ref{fig:planner}. We describe the resultant structure shown as the ``search tree" and each node in the tree represents a set of 3-tuple of (question, answer, reward). We refer to the initial query as a ``root node."}
\label{fig:search_tree}
\end{figure*}

\subsection{LLMs for Chemistry}

LLMs for chemistry can be divided into two categories: domain-specific models and adapted general-purpose models. Multimodal, domain-specific molecule-language models have recently emerged to target a number of problems in the chemistry domain \cite{edwards2021text2mol, vall2021bioassayclr, zeng2022deep, xu2022protranslator, su2022molecular, edwards-etal-2022-translation}. For brevity, we discuss them further in Appendix \ref{sec:mm_llm4chem}. On the other hand, general-domain LLMs such as GPT-4 \cite{openai2023gpt4} have been adopted in an agent-based approach to interface with chemistry-specific tools, allowing information gathering and hypothesis generation \cite{boiko2023emergent, bran2023chemcrow}. While this work is exciting, it differs from our approach, where we employ an overarching, domain-specific reward function to probe the LLM's knowledge, enabling stronger, simulation-backed guarantees about complex scientific reasoning. We note that our method builds upon \citet{sprueill2023monte}. Unlike their approach, and the approach of \citet{yao2023tree}, which use LLM-computed rewards, we integrate true computational chemistry-based rewards to guide the model. Further, we integrate a context-aware planner into our search algorithm to automatically guide the search.

\begin{algorithm}
	\SetAlgoLined
	\SetKw{Initialize}{Initialize}
	\textbf{Require:} LLM, initial prompt $P_{0}$, number of children to generate $N$, number of children to keep $M$, target depth $d$\\
    Initialize tree $T$ with nodes $P$ and edges $(P,a_j)$, scalar $\gamma$, stored values $p(P,a_j)$, and reward function $R$.\\
    $root(T) \leftarrow P_0$\\
    $P_{\mathrm{curr}} \leftarrow [root(T)]$\\
    $P_{\mathrm{next}} \leftarrow [~]$\\
    \For{$t=1,\dots,d$}{
        \For{$P_i\in P_{\mathrm{curr}}$}{
            $\mathcal{A}_i, p = \mathcal{P}(P_i,p)$$\rhd$ Get\ action set and priors\\
            \For{$a_j\in \argtopk{N}_{a_k\in\mathcal{A}_i}\left(p(a_j)\right)$}{
                $P_j^*\leftarrow a_j(P_i)$\ \ \ \ \ \ \ \ \ \ \ $\rhd$ Apply\ action $a_j$\\
                $T$.append$(P_j^*)$\\
                $P_{\mathrm{next}}$.append($P_j^*$)
                }
        $P_{\mathrm{curr}} \leftarrow \argtopk{M}_{P_j\in P_\mathrm{next}}(R(P_j))$ $\rhd$Calculate reward of LLM answer and downselect
        }
    }
    \Return $\arg\max_{P_j\in T}(R(P_j))$
	\caption{Description of the $\ourSystem$ framework.}
	\label{alg:heuristic_search}
\end{algorithm}
\section{System and Methods}
\label{sec:methods}

\begin{figure*}[h]
  \centering
  \includegraphics[width=0.7\textwidth]{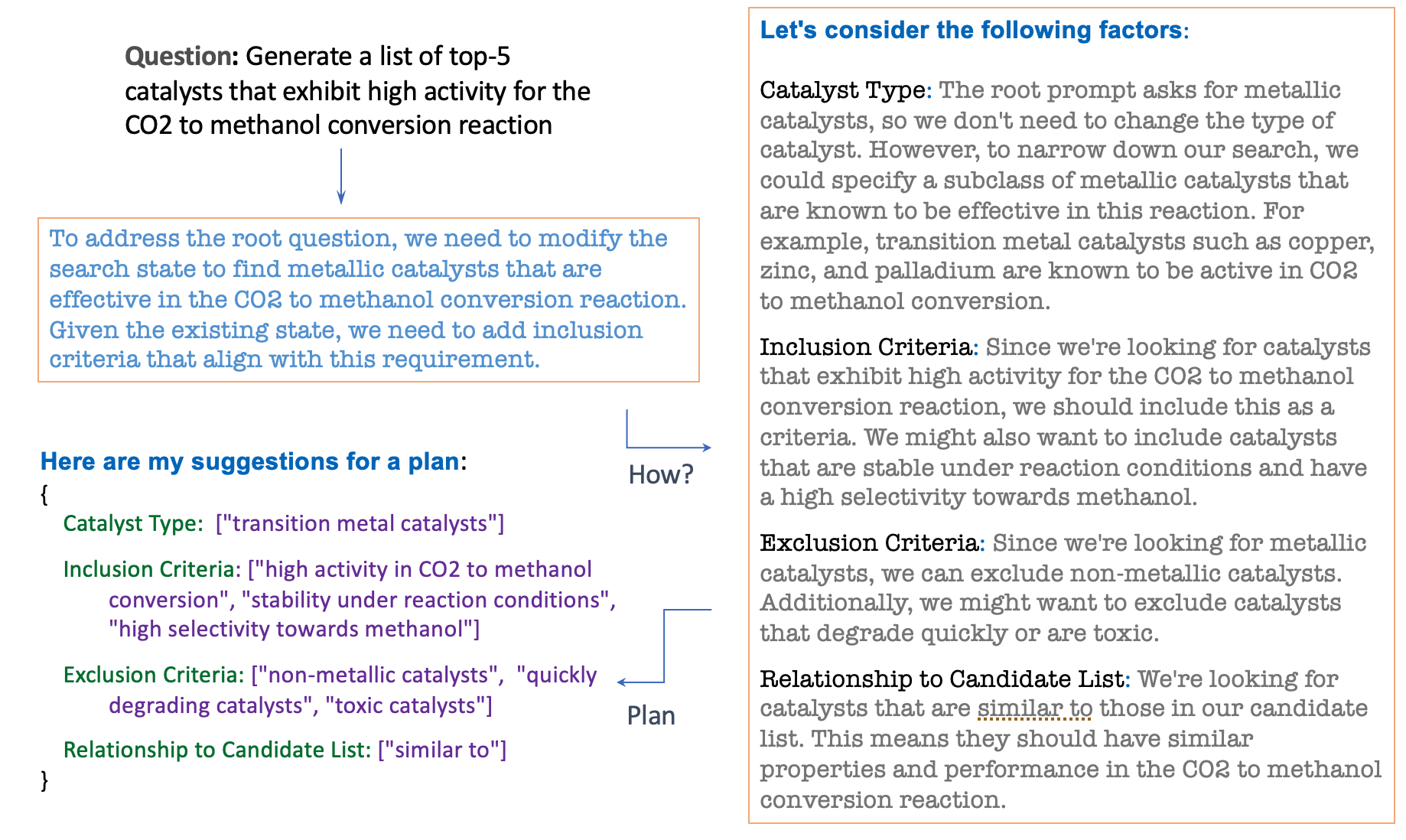}
\caption{Planner-guided search action generation (best viewed in color): Given a query state defined by a question (shown in top-left) and the set of corresponding answers, the LLM is used as an optimizer to generate a ``plan" for the next query. The LLM performs internal reasoning as shown in orange boxes. It accounts for the complete context from root query up to the current query node and generates a ``query plan" with the attributes ``catalyst type", ``inclusion criteria", ``exclusion criteria" and ``relationship to current candidate list".   
\vspace{-.2cm}
}
\label{fig:planner}
\end{figure*}

\begin{figure*}[!h]
  \centering
  \includegraphics[width=0.95\textwidth]{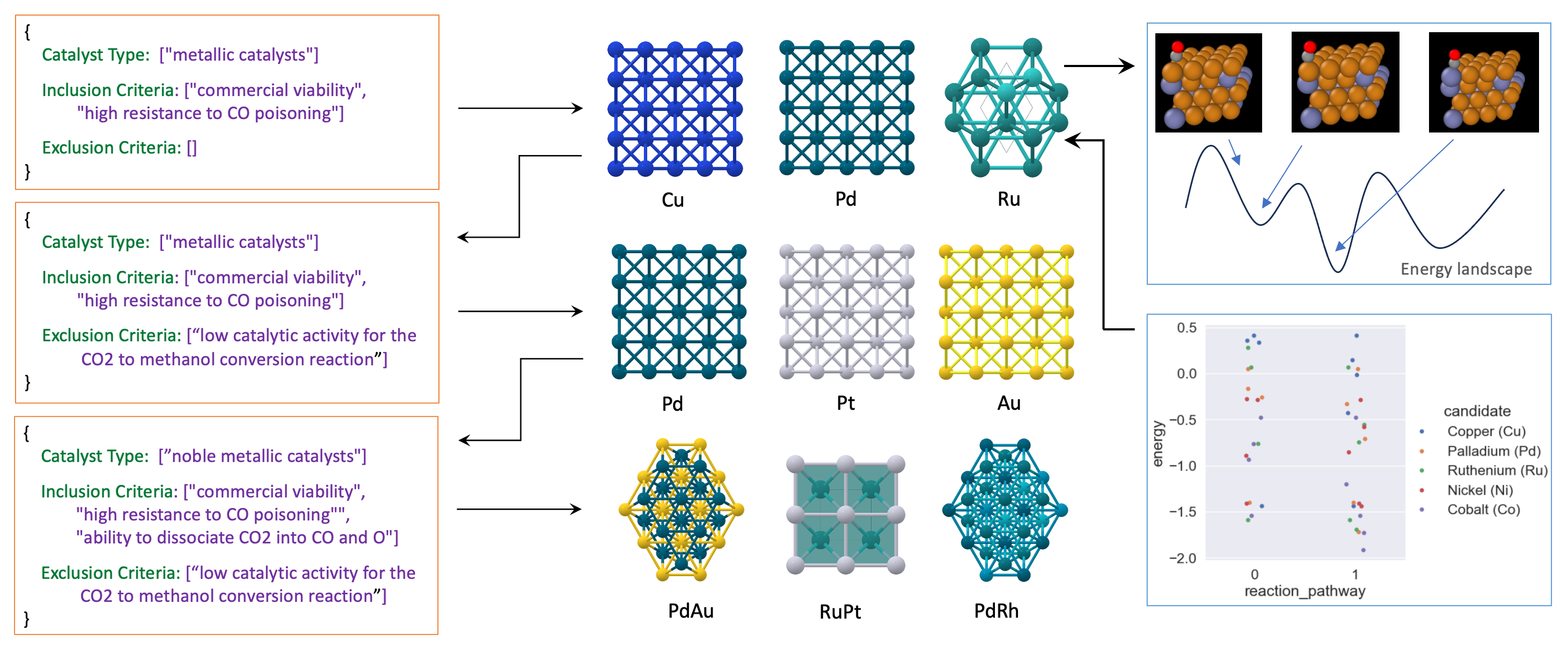}
\caption{Illustration of planner guided heuristic search (best viewed in color) described in section 3.2 below. Note the systematic expansion of the query plan in the orange boxes (left column).  The middle column shows illustration of 3D atomistic structures generated from chemical symbols.  Each 3D structure is processed by a reward function that involves geometry relaxation and potentially deriving approximations of energy barriers in reaction pathways (right column). Visualizations of materials acquired from the Materials project structure finder \cite{matproject}.}
\label{fig:reasoning_pathway}
\end{figure*}

This section elaborates algorithmic components of the architecture described in Algorithm  \ref{alg:heuristic_search}. It can be broadly divided into two components: 1) LLM-planned and guided heuristic search over chemical space and 2) quantum-chemical feedback from graph neural network (GNN) models trained from density functional theory (DFT) simulations.

\subsection{Heuristic Search}

The goal of our heuristic search is to answer a user-specified natural language query by systematically exploring candidates from different regions in the chemical space. Heuristic searches have been applied for general problem solving and reasoning tasks for LLMs \cite{yao2023tree}. Our work follows the catalysis-focused approach of \citet{sprueill2023monte}, where the original query (or prompt) and corresponding LLM answers are systematically modified by applying different screening criteria to iteratively contextualize the LLM prompts and answers into a narrower region of the chemical space. This process is illustrated in Figure \ref{fig:search_tree}, where the addition of chemical descriptors change the catalysts that are recommended. Since the number of branching pathways could hypothetically expand exponentially, we employ a domain-specific reward function to prune candidate catalysts that don't show promise of high catalytic activity. 



More formally, our goal is to search through chemical descriptors and design constraints to determine the optimal prompt, which leads the LLM to return the best candidate catalysts for a catalysis related query. Starting with a general prompt $P_0$, we use a set of actions to modify the prompt to improve the LLM output with respect to a reward function, $R$.  Notably, $\ourSystem$\text{-Planner} generates its own action space $\mathcal{A}$ (Figure \ref{fig:planner}).


\textsc{Definitions}: We define the \textsl{Search Tree} as a hierarchical tree consisting of (prompt, answer, and reward) nodes. Each node in this tree represents a state in the search space. Nodes are linked if an action $a \in \mathcal{A}$ modifies the prompt from one node to the next. We denote a path from root to leaf node a \textsl{Reasoning Pathway}. 

Following \citet{sprueill2023monte}, each node contains a template LLM prompt and an internal structured representation that provides additional context to the template prompt. This internal representation consists of 1) a natural language question, 2) an inclusion-exclusion list that includes or excludes specific chemical descriptors for target catalysts and 3) a relational operator that describes how the search can be shifted from the previous query's candidate catalysts to a different region in chemical space (i.e. similar-to or different-from the previous candidate catalysts). Starting the search with the root node, the search algorithm expands each node into a set of children nodes with a set of actions $\mathcal{A}$, which modify the internal representation of each node. The LLM then answers the modified prompt in each node, providing a set of candidate catalysts, and each candidate is scored using a reward function.  Each layer of the search tree is pruned using a beam search algorithm a beam search algorithm \cite{rubin1977locus}, leaving only those nodes with the highest rewards. Finally, when the maximum search depth is reached, we select the node in the tree with the highest reward as the best answer to the initial prompt.



\subsection{Planner-Guided Search}

Our planner component is responsible for systematically expanding the search by contextually determining viable actions. Specifically, the planner selects actions based on the current context of the search, using vocabulary relevant to the current catalyst candidates. This contextual grounding automatically constrains the search direction in a scientifically coherent way. A technical description of the planner prompt is given in Appendix \ref{app:sec:planner prompt}.

Consider any node in the search tree where the planner generates the actions for a given node (shown in the orange boxes of Figure \ref{fig:planner}). Next, we execute the actions, creating several child nodes, and retrieve a set of top-$k$ candidate catalysts from the LLM (such as Cu, Pd etc.). As shown in figure \ref{fig:reasoning_pathway}, each of these candidates are then transformed into a 3D atomistic representation and evaluated by a reward function, which approximates catalytic activity (see sections 3.3 and 3.4 for details). All nodes at a given depth in the search tree are collected and only a subset of nodes are chosen for further search in the next iteration, filtered by highest reward. The process repeats iteratively (shown via rows in above diagram) until the maximum tree depth is reached.




Overall, by leveraging the LLM to contextually expand the tree search, $\ourSystem$-Planner explores candidate catalysts in a more interpretable, scientifically-grounded reasoning pathways. See section \ref{sec:planner_trace} for a complete trace of a planner guided search, using GPT-4.

\begin{figure}[h]
  \centering
  \includegraphics[width=0.45\textwidth]{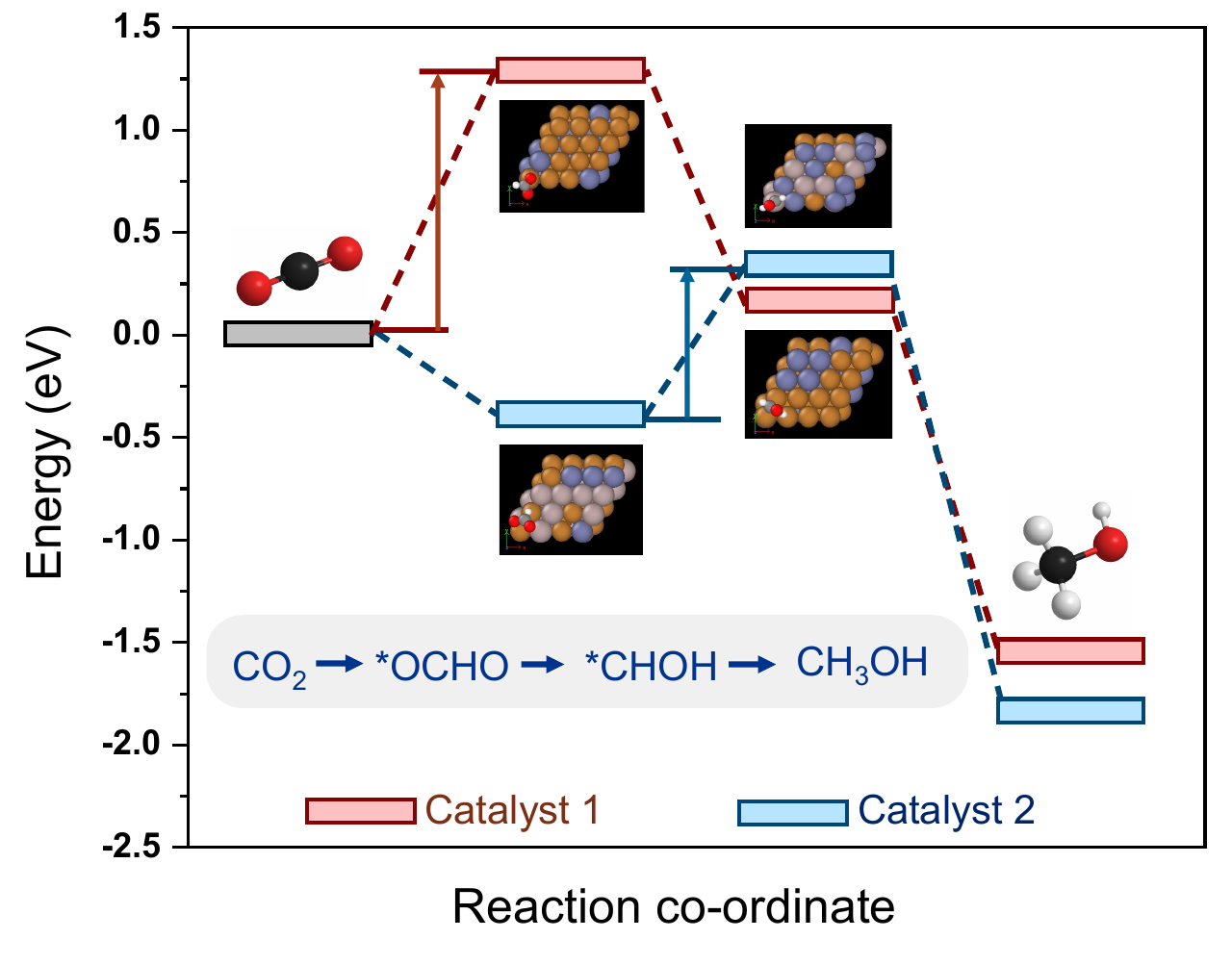}
\vspace{-.5cm}

\caption{
An illustration of reaction pathways corresponding to the conversion of \ce{CO2} to methanol for two different catalysts (red and blue bars shows the energy of different intermediates). The energy barrier (difference between the lower energy state and the higher energy state, described as hill climbing in the text) associated with these catalysts is shown by the red and blue arrows. For ease of comparison between the catalysts, the energies have been shifted such that both \ce{CO2} states have an adsorption energy of 0 eV.
}
\vspace{-.3cm}
\label{fig:rxn_reward}
\end{figure}

\subsection{Reward via Structure Optimization and Energy Prediction}

Each reward function returns a real number as a measure of the catalyst's goodness (higher is better) for a given input question. In this work, we implement two reward functions with different levels of complexity. The first reward function targets catalysts for the adsorption of particular chemical species, while the latter targets catalysts with higher approximate reactivity for certain reactions.

\textbf{Adsorption Energy-Based Reward}: To target catalysts for the adsorption of specific chemical species, we use an adsorption energy-based reward, which returns the adsorption energy of the most stable binding configuration of the catalyst as the reward. The computation begins with translating the symbolic representation of the catalysts (such as ``Platinum'') and adsorbates (e.g., ``*CO'') into a 3D atomistic structure (Fig. \ref{fig:reasoning_pathway} right). The stability and energy of a catalyst's atomic structure directly impacts its catalytic activity and selectivity. Therefore, we compute the most stable configuration for the catalyst-adsorbate pair and use its adsorption energy as a measure of the reward (Section \ref{app:sec:Relaxation Method}). The optimization process, also known as the \textsl{relaxation process}, iteratively relaxes the atomic positions of the 3D structure until an energy minimum is found. A GNN \cite{gasteiger2021gemnet} is then used to calculate the adsorption energy from this state.


\textbf{Reaction Pathway-Based Reward}: This function measures the goodness of a catalyst considering multiple reaction pathways and intermediate steps. We initially obtained 5 possible reaction pathways from the LLM for the \ce{CO2} to methanol/ethanol conversion reactions, as lists of chemical formulas. Since since some of the pathways were redundant, we manually narrowed down the responses down to 2 reaction pathways for each application. These prompts were not re-evaluated for each tree search.

Given each reaction pathway, our reward function computes the adsorption energies for every intermediate step. Figure \ref{fig:rxn_reward} shows two instances of the same reaction pathway for two different catalysts. As the figure shows, proceeding from one reaction step to another requires different amounts of energy, which depends on the catalyst and adsorbate interaction. Intuitively, moving from a lower-energy state to higher-one can be viewed as a ``hill-climbing'' in the energy landscape (indicated by the red and blue arrow for two different catalysts) and we formulate a function that would assign the highest reward to pathways with the smallest hills to climb (Eq. \ref{eqn:rxn_pathway_reward}). $\mathrm{ads}_t$ is the intermediate at step $t$ of the reaction and $E_{\mathrm{ads}_t}$ is the adsorption energy of $\mathrm{ads}_t$ on some catalyst. The overall reaction-based reward function for the top-$k$ catalysts is the average of $r(c)$, where $c$ in one of the top-$k$ catalysts and $r$ calculates the smallest maximum energy jump over all paths,
    
\vspace{-0.5cm}
\begin{equation}
\label{eqn:rxn_pathway_reward}
    r(c) = -\min_{p\in \mathrm{Paths}}\left(\max_{\mathrm{ads}_{t}\in p} \left(E_{\mathrm{ads}_t} - E_{\mathrm{ads}_{t-1}}\right)\right).
\end{equation}





\section{Experiments}

\begin{table*}
    \centering
    \begin{tabular}{c|cc|cc|cc}
         & \multicolumn{2}{|c|}{\textbf{OpenCatalyst}} & \multicolumn{2}{|c|}{\textbf{BioFuels}} & \multicolumn{2}{|c}{\textbf{\ce{CO2}-Conversion}} \\
         \hline
         & GPT-4 & GPT-3.5 & GPT-4 & GPT-3.5 & GPT-4 & GPT-3.5 \\
         \hline
        Chain-of-Thought & 0.37 & 0.66 & 2.08 & 2.10 & -0.62 & -0.54 \\
        Self Consistency & 0.73 & 0.76 & 2.08 & 2.12 & -0.54 & -0.36 \\
        \hline
        $\ourSystem$-Expert & 1.90 & 2.11 & 3.90 & 3.79 & 0.45 & \textbf{0.78} \\
        $\ourSystem$-Planner & \textbf{2.36} & 2.16 & \textbf{4.15} & 3.29 & 0.01 & 0.49 \\
        \hline
    \end{tabular}
    \caption{Final reward values of best recommended catalyst for each search variant. Our $\ourSystem$ methods both significantly outperforms the GPT-4 baseline. Larger numbers are better and reflect GNN-predicted adsorption or reaction pathway-based rewards.}
    \label{tab:rewards}
\end{table*}


We conduct an experiment to evaluate if $\ourSystem$, using an LLM-guided heuristic search with quantum-chemistry feedback, can discover more novel and effective catalysts than state-of-the-art LLMs alone. Our experiments focus on three key research questions for enabling such a system. 
\begin{enumerate}[align=left,start=3]
    \item[RQ1.] \textbf{Quantification of performance improvement}: Does heuristic-search guided by quantum-chemical feedback produce better catalyst candidates over querying state-of-the-art LLMs?  
    \item[RQ2.] \textbf{Characterization of key components}: What are the key parameters that control the computational complexity-system performance trade-off?
    \item[RQ3.] \textbf{LLM Hypothesis testing}: How do we verify hypotheses generated by $\ourSystem$ using domain knowledge?  Which areas need further attention to make $\ourSystem$'s computational screening more accurate and interpretable?
\end{enumerate}

\subsection{Experimental setup}


\textbf{Dataset} We conduct our experiments on an augmented version of a chemistry-focused reasoning query benchmark, originally proposed in \cite{sprueill2023monte}, containing 145 queries split between 3 general categories: OpenCatalyst, BioFuels, and \ce{CO2}-Fuel. We adopt queries from \citet{sprueill2023monte} for the first two categories and further enhance the dataset with the \ce{CO2}-Fuel subset. First, the OpenCatalyst dataset is compiled from the set of adsorbates from the Open Catalyst Project 2020 dataset \cite{chanussot2010open, zitnick2020introduction}; it requires suggesting catalysts which each adsorbate strongly binds to (86 queries). Second, the BioFuels dataset is targeted at catalyst discovery for biofuel development (39 queries). These queries have been modified to target metallic catalysts, which is necessary for our reward calculation. Finally, we specifically target the conversion of \ce{CO2} to methanol and ethanol (20 queries), a platform molecule, which can be used to produce fuels and chemicals for achieving net-zero carbon emissions \cite{ling2023transition, mondal2021methanol}.  See section \ref{sec:datasets} for details on the queries in these datasets



\textbf{System Implementation} The LLMs used in our experimental setup included OpenAI GPT-3.5 and GPT-4. Although we initially benchmarked $\ourSystem$ with LLama2 \cite{touvron2023llama}, we found that its instruction-following capabilities in this domain were too limited to allow an evaluation.

For our GNN reward model, we utilized the GemNet-dT model \cite{gasteiger2021gemnet} from the Open Catalyst Project. Runtime configurations and inference scaling performance for each of these models are provided in section \ref{app:sec:Scaling of LLM and GNN inference}. Inferences for both OpenAI models were executed in parallel using asynchronous execution features. The GNN inferences were run on a single DGX2/V100 or A100 GPU.


\section{Towards Explainable Reasoning from Chemical Feedback}
\subsection{Reasoning Approaches}
\label{sec:comparing reasoning approaches}
We evaluate two different variations of $\ourSystem$. $\ourSystem$-Expert is an implementation in which the action space is defined by catalysis experts. These actions (relation-operators and descriptors) are:
\begin{enumerate}[itemsep=0pt,parsep=0pt,topsep=0pt,partopsep=0pt]
    \item \textbf{Inclusion criteria}: high activity, high selectivity, low cost, novelty, low toxicity, high binding energy, high conversion, high availability.
    \item \textbf{Exclusion criteria}: low activity, low stability, low selectivity, low binding energy, high cost, high toxicity, low dispersion, low porosity, high scarcity, low conversion.
    \item \textbf{Catalyst type}: metallic catalysts, monometallic catalysts, bimetallic catalysts, trimetallic catalysts.
    \item \textbf{Relationship to previous candidate set}: include elements that are different from, include elements similar to, introduce new elements to, include elements from.
\end{enumerate}
These actions are sampled with uniform probability, without using the same criteria twice.
On the contrary, $\ourSystem$-Planner uses LLM-suggested actions for expanding the search space which do not require any expert specification. As such, $\ourSystem$-Planner's actions can adapt to the changing context of the search, whereas $\ourSystem$-Expert's action space remains static.

\textbf{Better LLM Translates into Search Efficiency} As Table \ref{tab:rewards} shows, both implementations of $\ourSystem$ significantly outperforms the GPT-4 baseline. This is also visualized by Figure \ref{fig:reward_scatter_plots} (section \ref{sec:Distribution of Reward Functions}). Specifically, $\ourSystem$-Planner coupled with GPT-4 performs best for the OpenCatalyst and Biofuels query categories, whereas $\ourSystem$-Expert performs best for \ce{CO2}-Conversion queries. As noted in Section \ref{sec:case_study} and shown in Table \ref{tab:case_study_lit}, the top-1 prediction of $\ourSystem$-Expert has high similarity with the current commercial catalyst for methanol synthesis \cite{etim2020improving}.  We also computed the average depth of the node containing the best answer for both variants of $\ourSystem$ (Table \ref{tab:search depth}). Lower average depth would indicate that our system is finding the best results more efficiently, and we observe that using GPT-4 leads to a reduction of 11.28\% in the average search depth. The impact is more pronounced for $\ourSystem$-Expert than $\ourSystem$-Planner, which already obtains performance boost through the algorithmic contribution of planning.

\textbf{LLM's Alignment with Reward function is key} The strong performance of the $\ourSystem$-Expert on \ce{CO2}-Conversion queries is noteworthy, especially considering it is based on GPT-3.5-turbo. We hypothesize that the performance is related to a complex reward function. For queries associated with the adsorption energy based reward function (OpenCatalyst and Biofuels) the LLM's notion of a good catalyst typically aligns with lower adsorption energy (higher reward) profiles. Therefore, the planner effectively uses the LLM as an optimizing function for searching towards energetically favorable catalysts. However, the LLM's notion of a good catalyst may not directly align with the complex reaction pathway-based reward function associated with \ce{CO2} conversion.  In general, it suggests that fine-tuning the LLM using a methodology similar to RLHF \cite{ouyang2022training} may be a promising path for downstream tasks with complex reward functions.


\subsection{Performance Characterization of Key Components}

$\ourSystem$ performs a large number of LLM and GNN inferences that influence both its performance and throughput. The following factors control the computational complexity of the overall execution.

\textbf{LLM Inference and Tree Search} The runtime of $\ourSystem$ is $O(N_{tree})$, where $N_{tree}$ is the number of nodes in the search tree. $N_{tree}$ increases exponentially with the maximum depth (set to 5) and the branching factor for expanding each node in the search tree, denoted as $N_{actions}$ (set to 8). We use a beam width parameter (set to 6) to control the number of nodes that are expanded in each iteration of the search.
$\ourSystem$ performs LLM inferences for each node in the search tree to 1) plan actions, 2) suggest candidate catalysts, and 3) transform LLM answers into atomistic structures. It is important to note the strong intra-node and inter-node dependencies between these LLM queries, as the output from one query is fed into the next ones. This dependency exists both within a node and to its children in the search tree. Therefore, robustness of LLM instruction-following behavior, specifically the ability to return answers in a consistent format, is a critical factor for an effective system. Overall, $\ourSystem$-Expert and $\ourSystem$-Planner executes $2 N_{tree}$ and $3 N_{tree}$ LLM inferences respectively. 

To execute the entire benchmark of 145 queries with $N_{max} = 300$ requires execution of 28,000-42,000 LLM inferences, which makes the the scalability and throughput of a generative LLM a critical factor for successful experimentation. By default, we batch LLM inferences in sizes of 48 and execute these inferences asynchronously for scalability. These two factors, namely the robustness of instruction-following and high-throughput execution of batched queries restricted our experiments to only GPT-3.5 and GPT-4 models. Our experiments with LLama2 were both affected by frequent spurious answers and relatively low throughput on A100 systems. We recognize the importance of the development and benchmarking of open-source models, but this particular topic should be the subject of future work.

\textbf{GNN Inference and Reward Estimation} GNN inferences play an equally critical role for $\ourSystem$. A single execution of the adsorption energy-based reward function requires multiple inferences. Given a string-based representation of a catalyst, we first randomly initialize a 3D atomistic representation and oversample the structural conformation to generate $N_{structs}$ samples (see Appendix \ref{app:sec:Relaxation Method}). Next, relax each of these $N_{structs}$ structures (set to 16 by default) for a maximum specified iteration limit of $N_{relax}$ (set to 64 by default), or until $\mathbf{f}_{\mathrm{max}}$. Therefore, each execution of each adsorption energy-based reward function requires $N_{structs}*N_{relax}$ GNN inferences. For scalability, we group the GNN inferences using a batch size of 40. In reaction pathway-based reward computations, we perform this computation for every intermediate state in each pathway. Therefore, with $N_{pathway}$ denoting the number of pathways and $N_{rstep}$ denoting the average number of intermediate steps per pathway, a single reward estimation involves $N_{structs}*N_{relax}*N_{pathway}*N_{rstep}$ GNN inferences. Given our \ce{CO2}-fuel conversion queries which involve 2 reaction pathways comprising 4-5 steps, we execute 9,216 GNN inferences for every reaction-pathway based reward. On a single A100 or V100 GPU, batches can be evaluated in approximately 0.5 seconds.

\subsection{Considering DFT-Based Rewards}

A critical component for a reasoning system is accurate feedback. For this reason, the use of DFT-based rewards would be the gold standard reward calculations. However, due to their significant computational cost, we are forced to employ GNNs for high-throughput evaluation of catalyst candidates \cite{zitnick2020introduction}. We evaluated our top GNN results using DFT simulations; overall, we found that DFT calculation-based rewards diverge in some cases significantly from GNN predictions (Table \ref{tab:dft_results}). Here, we briefly examine the divergence between SOTA GNN models and ground truth DFT simulations. 

One limitation of the GNN likely arises from the conversion of text-based catalyst recommendations from the LLM to 3D atomistic structures for the GNN calculation. Our structure generation method (see Section \ref{app:sec:Relaxation Method}) is limited to pre-defined lattice structures, (e.g., face centered cubic (FCC), body centered cubic (FCC), and hexagonal close packed (HCP)). While these lattice structures effectively describe mono-metallic structures, multimetallic compounds may exhibit more complex lattice structures. Thus, our assumptions of reference structures may not adequately reflect realistic catalyst structures. Furthermore, our multimetallic bulk structures were generated by randomly placing elements throughout the bulk, creating additional complexity due to the emergence of defects in the bulk structures.

While the GNN can predict energy for any given structures, the challenges mentioned above may lead to erroneous DFT calculations, which makes validation of our predicted catalysts difficult. Therefore, future research should explore novel methods to convert textual representations of catalysts into realistic 3D atomistic structures.



\subsection{\ce{CO_2} Hydrogenation: A Case Study} \label{sec:case_study}

To evaluate the real-world efficacy of $\ourSystem$, we performed a literature evaluation of the top-5 predicted metallic catalysts for the conversion of \ce{CO_2} to methanol. These predictions from GPT-4, $\ourSystem$-Expert, and $\ourSystem$-Planner models are shown in Table \ref{tab:case_study_lit}. We find that the $\ourSystem$-Expert predictions include elements that make up the current commercial catalysts for methanol production Cu/ZnO/\ce{Al_2O_3} \cite{etim2020improving}. Cu, ZnO, and Cu-Zn sites were considered to generate the active sites for high methanol selectivity \cite{etim2020improving, TISSERAUD2015533}. 

On the other hand, $\ourSystem$-Planner predicted bimetallic precious metal alloys, which are well-known to catalyze hydrogenation reactions \cite{TAWALBEH2023103217}. The ensemble effect \cite{doi:10.1021/jacs.9b05766} in Pd-Au electrocatalysts was considered as key factor in achieving high activity for \ce{CO_2} reduction to methanol since Pd, Pt, Rh and Ru alone can shift the selectivity to methane (\ce{CH_4}) by uncontrolled hydrogenation of \ce{CO_2} \cite{C3CS60395D}. Alloying allows for improved methanol selectivity compared to the performance of single metal catalysts \cite{TAWALBEH2023103217, etim2020improving, TISSERAUD2015533} that were predicted by GPT-4 and some by $\ourSystem$-Expert. Since both $\ourSystem$-Expert and $\ourSystem$-Planner predicted alloys, we consider their recommendations to have higher likelihoods of being successful catalysts.  

\section{Conclusion}

We introduce a multi-modal framework unifying linguistic reasoning enabled by generative LLMs with atomistic structure based rewards, built on principles in catalysis and quantum chemistry. We demonstrate that such integration can enable the recommendations of catalysts that are critical for the development of energy efficient chemical conversion processes that combat climate change. Namely, $\ourSystem$ was able to recommend CuAlZn for the conversion of \ce{CO2} to methanol, the current commercially viable catalyst.

Our methodology makes use of instruction-following foundational LLMs, GPT-3.5 and GPT-4, avoiding the need for expensive finetuning. Additionally, our application of automatic planning methods removes the need to hard-code domain specific knowledge into the search method. So long as one can define a systematic search space and a relevant domain specific reward function, one could adapt our method to other scientific applications in biology, chemistry, or materials science. While we recognize the importance of accessible, open-source LLMs, our experiments with Llama-2 revealed that such models may not currently possess the instruction following capabilities or scalability required for our heuristic search. Therefore, the development of open-source LLMs for scientific reasoning remains the subject of future work.

Finally, $\ourSystem$ presents a new way to carry out computational investigations of catalyst materials with DFT surrogate models. Our application of a GNN based reward function grounds the LLM in reality, guiding the search towards catalysts with higher activity. Still, a limitation of $\ourSystem$ is the reliance on existing reference structures, which may not accurately reflect real catalyst structures, providing a challenge to verification of results with DFT. This motivates the future development of more robust methods to convert textual representations of materials into 3D atomistic structures.


\section*{Acknowledgements}

This work is supported in part by the PNNL seed
Laboratory Directed Research and Development
(LDRD) program, Generative AI for Science, Energy and Security Lab-Wide Laboratory Directed Research and Development Investment, and the Microsoft Accelerating Foundation Models Research program.

This work is also in part based upon work supported by the Molecule Maker Lab Institute: an AI research institute program supported by NSF under award No. 2019897. The views and conclusions contained herein are those of the authors and should not be interpreted as necessarily representing the official policies, either expressed or implied, of the U.S. Government. The U.S. Government is authorized to reproduce and distribute reprints for governmental purposes notwithstanding any copyright annotation therein.

PNNL acknowledges the Microsoft Azure Quantum Elements team for their assistance with this study and their contributions to this publication.

\section*{Impact Statement}

This paper presents work whose goal is to advance the field of Machine Learning for the chemical and material sciences. It involves the use of a large language model (LLM), which should be treated as a black-box. LLMs are known to ``hallucinate''---produce erroneous or fabricated text---and may plagiarise other scientific work. The goal of this paper is to incorporate a domain specific reward function to mitigate hallucinations by the LLM by validating their answers with computational chemistry feedback. Still, scientific results produced by LLMs should still be stringently verified by domain-experts to avoid false claims. The primary focus of this work is to discuss the strengths and weaknesses of the AI system proposed herein as a tool to accelerate scientific discovery, not necessarily to claim the discovery of a novel catalyst.






\bibliography{iclr2024-latex/anthology,iclr2024-latex/main}
\bibliographystyle{iclr2024-latex/icml2024}

\newpage

\clearpage
\appendix

\section{GNN Driven Adsorption Energy Calculation}

Here, we detail our methodology for GNN-driven adsorption energy calculations. Given a linguistic catalyst recommendation from the LLM, an adsorption energy is produced in a three step process, catalyst generation, adsorbate+catalyst sampling, relaxation.

First, the linguistic representation of a catalyst from the language model must be translated into a formate which is computationally digestible. For instance, a simple catalyst, like pure Copper Zinc (CuZn), can be expressed in several different ways:
\begin{itemize}
    \item Copper-Zinc
    \item Copper alloyed with Zinc
    \item CuZn
    \item Copper/Zinc-oxide (catlyst not supported by our method)
    \item Cu-Zn.
\end{itemize}
Attempting to capture all possible disambiguation of catlyst is intractable for large tree searches, especially with less-well behaved models such as GPT-3.5-turbo. It also becomes more difficult to decide which catalysts to penalize as containing non-metals, such as Copper/Zinc-oxide. Therefore, we prompt the LLM to parse out the correct chemical symbols into digestible lists of chemical symbols (see Appendix \ref{app:sec:syms prompt} for the prompt details). We would expect the above phrases to be parsed into:
\begin{itemize}
    \item[\labelitemi] [``Cu'', ``Zn'']
    \item[\labelitemi] [``Cu'', ``Zn'']
    \item[\labelitemi] [``Cu'', ``Zn'']
    \item[\labelitemi] [``Cu'', ``Zn'', ``O''] (catlyst not supported by our method)
    \item[\labelitemi] [``Cu'', ``Zn''].
\end{itemize}

These examples are much easier to parse out, since we expect to see only chemical symbols. Additionally, the ``O'' can easily be parsed out to return a penalty value. With these lists of chemical symbols we can use a set of rules to generate the associated 3D structures.

In the second step, the 3D structure of the catalyst is inferred from known reference structures, as specified by the Atomic Simulation Environment (ASE) package \cite{ase-paper}. For the purposes of this paper, we restrict our focus only to catalyst whose reference structures are in face-centered cubic (FCC), body-centered cubic (BCC), or hexagonal close-packed (HCP) lattice structures. Catalysts with other reference structures are skipped and assigned a negative reward value. However, many important metals for catalysis have these three lattice structures. If the recommended catalyst is a bimetallic or trimetallic catalyst (composed of two or three elements), we randomly introduce atoms from the secondary and tertiary elements into the structure. For bimetallic compounds, AB, we assume a 2:1 ratio of A to B, while trimetallic catalysts have a 1:1:1 ratio. Catalysts with more than 3 elements return a penalty value. The lattice structure is always inferred by the first species listed by the LLM. For instance CuZn would have the structure of copper, with some atoms randomly replaced with Zn in a 2:1 ratio. To promote the stability of the catalysts, we sample 16 possible configurations and chose only the structure with the lowest energy, as determined by the GNN, to move on to the rest of the GNN calculation.

The third step of the process is the sampling of adsorbate+catalyst configurations. the goal is to sample different locations and orientations for an adsorbate to bind to the surface of the catalyst, to ensure optimal binding is observed. We determine binding sites, use the placement sites specified by ASE for each of the three lattice configuration. Then, for an adsorbate, which we denote generically as, \mbox{*}XYZ, we place the atom marked by the \mbox{*} at the binding location at a height 1.87 \AA above the surface. Then, to sample orientations of the adsorbate, we rotate the molecule a random angle up to 15 degrees in the x-axis (axis parallel to the surface), then a random angle up to 360 degrees around the z axis (axis perpendicular to the surface). Finally, since some adsorbate molecules may bind differently to certain elements in multimetallic catalysts, we randomly replace the binding site of the catalyst with each of the available metallic elements, with equal probability. This gives fair change of the adsorbate binding to any of the given catalyst species. We take 16 initial adsorbate+catalyst configurations and relax each of them with the GNN to compute adsorption energies.

The fourth and final step is the the GNN driven relaxation. Given the set of atomic coordinates and atomic numbers of a structure, the GemNet-dT model \cite{gasteiger2021gemnet} returns an approximation to that structure's adsorption energy and forces, $E_{\mathrm{ads}}$ and $\mathbf{F_{\mathrm{max}}}$. Starting with the sampled adsorbate+catalyst configurations, the structures are relaxed in-batches of 40, using the L-BFGS from Pytorch \cite{ocp_dataset,NEURIPS2019_9015}. We stop the relaxation terminates when after 64 stems or when the magnitude of the maximum single atom force is below the early stopping threshold, $\mathbf{F_{\mathrm{max}}} < 0.05$ eV/\AA. Then, the minimum adsorption energy of the 16 samples is returned as the final adsorption energy for the adsorbate+catalyst configuration.

These four steps constitute the computational backbone of the reward calculation. For reaction based queries, the reward function is derived from these energies according to Equation \ref{eqn:rxn_pathway_reward}.


\label{app:sec:Relaxation Method}
	

\section{Dataset Design }
\label{sec:datasets}

We propose two task datasets related to catalyst design: the first is derived from the Open Catalyst (OC) Project \cite{zitnick2020introduction} and the second consists of complex reasoning queries designed by catalysis experts.  Our multi-disciplinary team involves researchers who actively work on designing new catalysts for bio-fuels development.

\subsection{Prompt Design}
\label{sec:prompt_and_action_definitions}
To apply $\ourSystem$ to catalyst discovery, we define use a prompt template which incorporates and overall query for the LLM to answer and we use the actions defined in \ref{sec:comparing reasoning approaches} to modify the fields of the templates to reason about the queries. Here, we discuss the s general prompt template, which is used for $\ourSystem$, CoT, and self-consistency, before diving into the construction of queries within the datasets.

Given a query, we use a prompt template which allows the systematic adjustment of an LLM's context for answering the query, essentially creating a reasoning process. The reasoning has four changeable variables, the ``catalyst type'', ``inclusion criteria'', ``exclusion criteria'', and ``candidate\_list\_statement''. Here, we describe each of these variables in detail.

First, the ``catalyst type'' is the broad category of catalysts to suggest (i.e. metallic catalysts, bimetallic catalysts, transition metal catalysts). Second, we provide a list of `inclusion criteria'' and ``exclusion criteria'' to provide additional context for which properties are more/less useful for suggesting a catalyst that answers the original query. Such examples could be ``include catalysts with low cost, high conversion, high Lewis-acidity'' or ``exclude catalysts which degrade quickly''. While the variables provide context to the LLM, in $\ourSystem$, we additionally include, in the prompt, the suggested catalysts from the previous answer as additional context. This motivates final variable, ``candidate list type,'' where the prompt LLM is instructed how to use the previous candidates in its evaluation. For instance, it can produce candidates that are ``similar to..'', ``different from...'', or ``have stronger affinity for oxygen than...'' the previous candidates.

The overall prompt template is as follows: 

{\tt \{query\}

\{include\_statement\}\{exclude\_statement\}
Provide scientific explanations for each of the \{catalyst\_label\}.
Finally, return a python list named final\_answer which contains the top-5 \{catalyst\_label\}.
\{candidate\_list\_statement\}

Take a deep breath and let's think step-by-step. Remember, you need to return a python list named final\_answer!}

The query in the prompt template is determined by our three datasets: OpenCatalyst, BioFuels,

\subsection{OpenCatalyst Dataset}

The Open Catalyst dataset \cite{zitnick2020introduction} is an extensive collection of (DFT) calculations for various adsorbate+catalyst, intended for training surrogate models for computational chemistry simulations related to catalysis. Here, we utilize this dataset to benchmark the ability of the \ourSystem to produce catalyst recommendations for the set of adsorbates given in the Open Catalyst 2020 dataset.

Each query consists of a single adsorbate from the Open Catalyst Project dataset (i.e. \ce{\mbox{*}OHCH3}, where \mbox{*} denotes the catalyst binding site). So, each prompt template takes the form {\tt Generate a list of top-5 \{catalyst\_label\} for the adsorption of \{adsorbate\}}, where {\tt catalyst\_label} begins with {\tt `metallic catalysts'}. 

\subsection{BioFuels Dataset}

The BioFuels dataset targets more complex adsorbates related to (a) the production of sustainable fuels from organic materials and (b), the reverse water gas shift reaction. The reactions and adsorbates targeted here are crucial to the production synthetic bio fuels with greater selectivity.\cite{canakci1999biodiesel, daza2016co, kattel2017tuning, artz2018sustainable, stolarczyk2018challenges, xu2018theoretical, mukhtar2022current} 

For the queries in (a), we benchmark the ability of \ourSystem 
 to recommend catalysts for the adsorption of specific adsorbates related to bio-fuel production. Such adsorbates are phenol(-ate), anisole, furfurol, methanol, methyl, ethanol, acetic acide, and acetate. Additionally, we apply a criterion of interest to \ourSystem, such as high/low selectivity, conversion, binding energy. Finally, we provide additional context by specifying a type of reaction we are interested in, hydrodeoxygenation or hydrogenation. The template for these queries is {\tt Generate a list of top-5 \{catalyst\_label\} can bind \{adsorbate\} in the \{reaction\_type\} reaction with \{additional\_constraint\}}, where catalyst\_label begins with {\tt `metallic catalysts'} while {\tt reaction\_type}, {\tt adsorbate}, and {\tt additional\_constraint} take the values described above. 

For the queries in (b), we benchmark the ability of \ourSystem to recommend catalysts for the production of \ce{CO} through the RWGS reaction,
\begin{align*}
    \ce{CO2 + H2 <=>[cat] CO + H2O}.
\end{align*}
There are several query variations for this reaction. First, we ask \ourSystem to recommend catalysts that strongly bind the reactants (\ce{CO2} and \ce{H2}) and weakly bind the product (\ce{CO}). The remainder of the questions relate to catalysts which demonstrate high convergence for the reaction. Then, we change the type of catalyst requested (metallic catalyst, bimetallic catalyst, bimetallic catalyst with noble metal and base metal). Finally, we optionally include an additional criterion for low cost catalysts. The final query template is {\tt Generate a list of top-5 \{catalyst\_label\} \{additional\_criterion\} for \{reaction\_question\}}.

\subsection{\ce{CO_2}-Conversion Dataset}

For the final set of queries, the \ce{CO_2}-conversion dataset, we propose a set of 20 queries to benchmark \ourSystem's ability to recommend catalysts for more complex reactions, the conversion of \ce{CO2} to methanol and ethanol (X-anol). Compared to the RWGS reaction, these reactions consist of pathways with different chemical intermediates. The language model will have to consider how the structure of the catalyst promotes the progression of the reaction of several chemical intermediates. Thus, queries involving the \ce{CO2} to X-anol conversion reactions use our reaction pathway based reward.

These query templates echo the same structure as the RWGS templates, {\tt Generate a list of top-5 \{catalyst\_label\} \{additional\_criterion\} for \{reaction\_question\}}.

\begin{table*} 
    \centering
     
    \begin{tabular}{cccc}
        Top-5 Catalysts & GPT-4 & $\ourSystem$-Expert & $\ourSystem$-Planner\\
        \hline
        1 & Pd & Cu-Zn & Pd-Au \\
        2 & Cu & Fe & Pt-Ru \\
        3 & Ru & Ni & Ru-Au \\
        4 & Rh & Co & Rh-Pd \\
        5 & Pt & Cu-Cr & Pt-Au \\
    \end{tabular}
    \caption{Final catalyst recommendations from $\ourSystem$ and GPT-4 (root of the tree) for the \ce{CO2}$\rightarrow$ methanol conversion reactions.}
    \label{tab:case_study_lit}
 \end{table*}

\subsection{Symbols Parsing Prompt}

One critical requirement of $\ourSystem$'s pipeline is the ability to a fuzzy, linguistic catalyst name into a list of symbols (see Appendix \ref{app:sec:Relaxation Method}). To accomplish this task, we use the following LLM prompt template:
{\tt Consider the following list of catalysts:

[\{cat\_0\}, \{cat\_1\}, ..., \{cat\_k\}].

For each catalyst, return the list of chemical symbols that make up the catalyst. If a catalyst does not have a chemical symbol, return None. If a catalyst is already a chemical formula, repeat the elements in the chemical formula.

Format your list as:

\{cat\_0\}: [list\_0]\\
\{cat\_1\}: [list\_1]\\
...\\
\{cat\_k\}: [list\_k]
}

Where {\tt cat\_i} is the {\tt i}th catalyst proposed by the LLM. The symbols lists can be parsed out from the answer by following the provided format in the prompt.

\label{app:sec:syms prompt}

\subsection{Planner Prompt}

Here, we detail how $\ourSystem$ determines the next set of catalyst descriptors by implementing a planner prompt. The prompt incorporates context from previous catalyst answers and the current search state to update the search states in a reasonable way. The search state corresponds to the 4 action types listed in section \ref{sec:comparing reasoning approaches}:
\begin{enumerate}[itemsep=0pt,parsep=0pt,topsep=0pt,partopsep=0pt]
    \item {\bf Catalyst Type}: the current catalyst type,
    \item {\bf Exclusion Criteria}: the list of catalyst design criteria to exclude,
    \item {\bf Inclusion Criteria}: The set of catalyst criteria to include,
    \item {\bf Relationship to previous candidate set}: How the new catalyst should relate to the previous catalysts (e.g. similar to, different from).
\end{enumerate}

\begin{figure}[h!]
  \centering
  \includegraphics[width=0.45\textwidth]{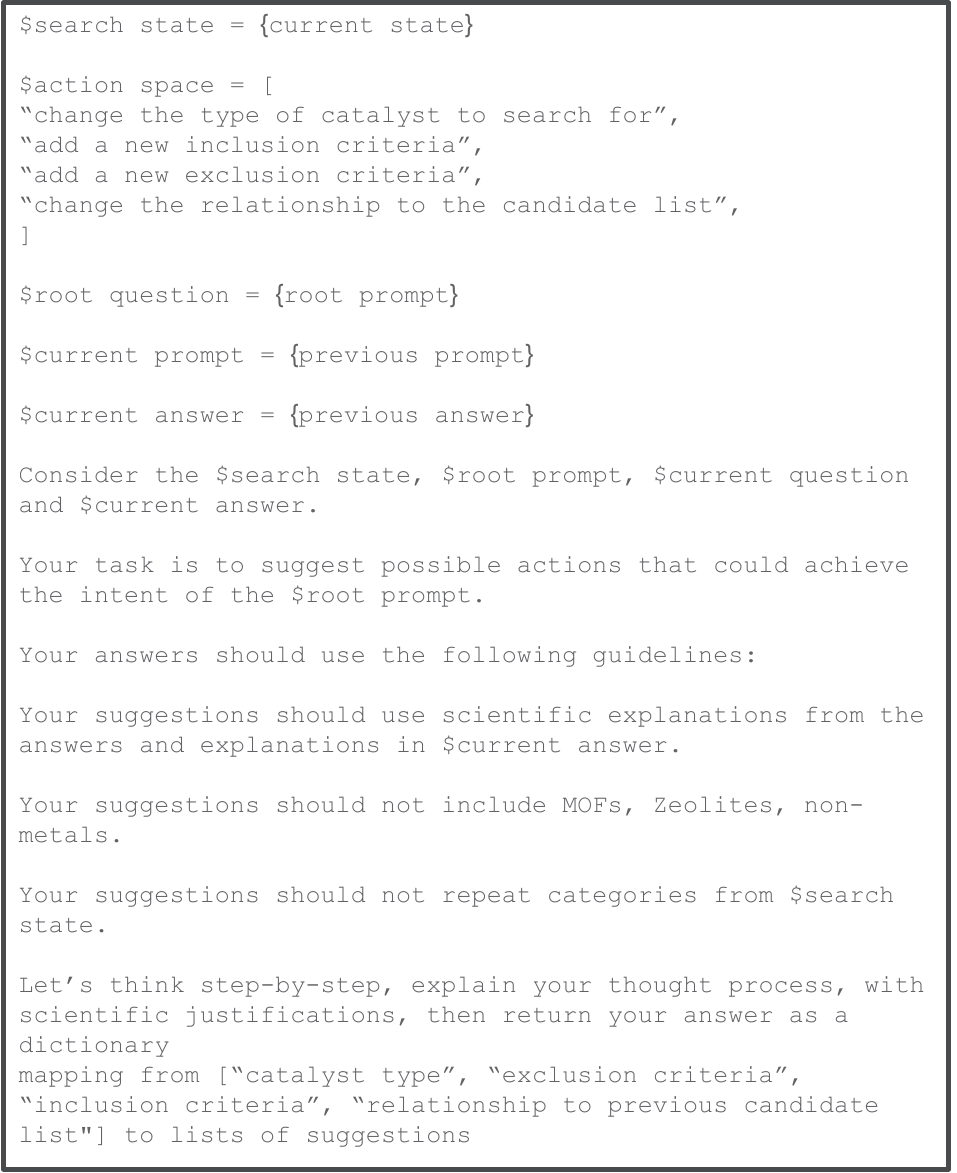}
\caption{The planner prompt template. The LLM is prompted to suggest actions which modify the four state variables. Context about the current search state is provided to the LLM by using ``variables,'' denoted with a \$ symbol.}
\label{fig:planner_prompt}
\end{figure}

$\ourSystem$ uses actions to modify these four state variables. Whereas $\ourSystem$-Expert uses a hand-defined set of actions  designed by catalysis experts, $\ourSystem$-Planner uses the LLM to determine the actions. To this end, we propose the following contextualized prompt template for our planner module, presented in figure \ref{fig:planner_prompt}.

In the planner template, the phrases denoted by {\tt\$} are treated as ``variables'' in the LLM's context window. They hold information about the following information:

\begin{enumerate}[itemsep=0pt,parsep=0pt,topsep=0pt,partopsep=0pt]
    \item {\tt search\_state}: a dictionary mapping each $\ourSystem$ state variable name to its current value,
    \item {\tt action\_space}: a description of the possible actions,
    \item {\tt root\_question}: the root question for $\ourSystem$ to answer. For example, ``What is a good catalyst for X reaction?''
    \item {\tt current\_prompt}: the previous catalyst prompt that produced the {\tt current\_answer}. It should be a modification of {\tt root\_question} with the previous actions chosen by $\ourSystem$,
    \item {\tt current\_answer}: the current catalyst recommendations and their scientific explanations, as returned by the LLM.
\end{enumerate}

\label{app:sec:planner prompt}

\section{Search Analysis}

\subsection{Parameter selection} 
\begin{figure}
    \centering
    \includegraphics[width=0.95\linewidth]{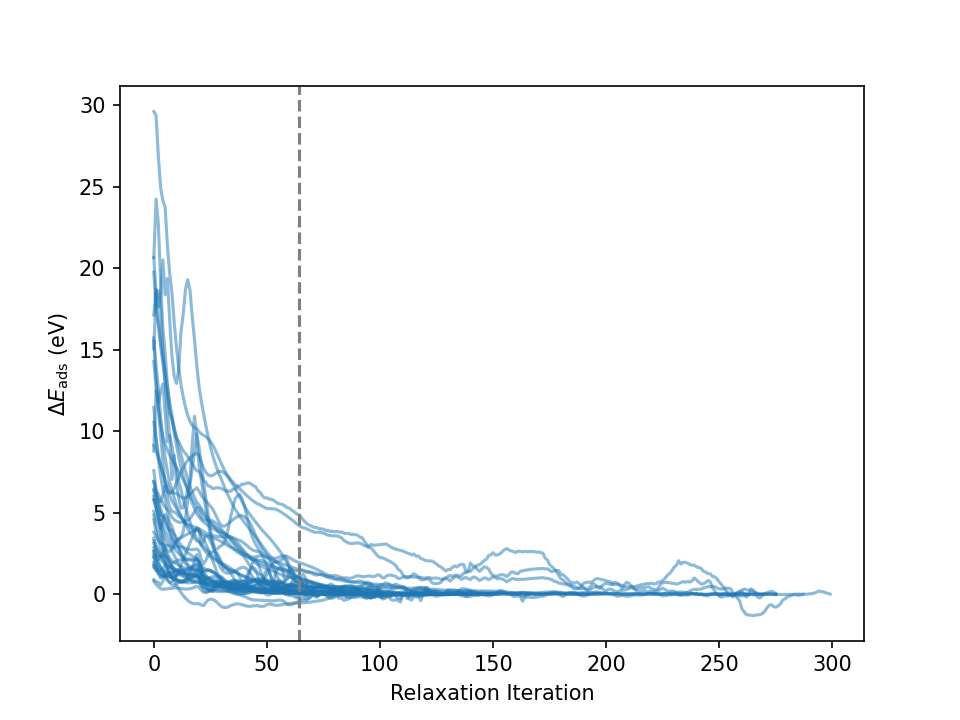}
    \caption{GNN energy versus relaxation iteration for 40 randomly chosen adsorbate+catalyst configurations. The difference between the current and final GNN energies are displayed to highlight the time-to-convergence for each relaxation. The majority of structures are nearly fully relaxed by 64 iterations.}
    \label{fig:gnn_convergence}
\end{figure}

Given the compute intensive nature of each $\ourSystem$ query, we carried out hyperparameter sweeps to determine the optimal number of actions ($N_{actions}$), as well as the maximum number of relaxation steps ($N_{relax}$). Figure \ref{fig:num_actions} shows how increasing $N_{actions}$ impacts the distribution of rewards, and we set it to a default value of 8 for all experiments.  Figure \ref{fig:gnn_convergence} shows convergence of the GNN for finding the optimal atomistic structure configuratio over 300 iterations for 40 random initial adsorbate+catalsyst configurations. Relaxations are terminated early if $|\mathbf{F_{\mathrm{max}}}| <0.05$ ev/\AA. At each step in the relaxation process, the difference between the current adsorption energy and the final adsorption energy is reported. The dashed gray line depicts the iterations cutoff for reward calculations in \ourSystem, 64 steps. Most iterations are nearly converged by 64 steps.

\subsection{Quantitative Validation with DFT}

To validate the adsorption energies being predicted by the GNN, density functional theory (DFT) calculations were performed for the top scoring candidates.
These calculations were carried out using the open-source \mbox{\emph{Quantum ESPRESSO}} package \cite{giannozzi2009quantum, giannozzi2017advanced} which implements DFT using a plane-wave basis set. Core electrons were replaced by pseudopotentials of  projector augmented-wave (PAW) type from the \emph{pslibrary} \cite{dal2014pseudopotentials}. The \emph{PBE} exchange-correlation functional was used \cite{perdew1996generalized}, and van der Waals effects were included using the DFT-D3 approach \cite{grimme2010consistent}. All of the calculations were managed and orchestrated using the AiiDA automation framework \cite{huber2020aiida}.

While the GNN predicts the adsorption energy directly, each DFT computation of the adsorption energy requires three calculations of the total energy:

\begin{equation}
    E_{adsorption}^{DFT} = E_{system}^{DFT} - ( E_{slab}^{DFT} + E_{adsorbate}^{DFT} )
\end{equation}

where $E_{system}^{DFT}$ is the DFT total energy for the complete system, $E_{slab}^{DFT}$ is the DFT total energy for the bare slab without any adsorbate, and $E_{slab}^{DFT}$  total DFT energy of the adsorbate molecule alone. 

To compute each of these energies, relaxations were initiated from the final relaxed coordinates coming from the GNN, removing the adsorbate or the slab if necessary. For the both the GNN relaxations and the DFT relaxations, the same atoms in the bulk layers were fixed, while those in the surface layer were allowed to relaxed. The cell vectors from the GNN structures already included sufficient vacuum above the slab and so were used directly for the corresponding DFT calculations.

\begin{table}[]
    \centering
    \footnotesize
\begin{tabular}{lllSS} \toprule
{Surface} & {Pathway}  & {Adsorbate} & \multicolumn{2}{c}{Adsorption Energy (eV)}        \\ \cline{4-5} 
          &            &            & {GNN}  & {DFT}  \\ \midrule
{CuZn}    & Methanol   & \ce{CO2}   & 0.384  & -0.066 \\
          &            & {CHOH}     & 0.552  & 5.951  \\
          &            & {OCHO}     & 0.577  & 5.836  \\
          &            & \ce{OHCH3} & -1.160 & 2.699  \\ \midrule
CuAlZn    & Methanol   & {CO$_2$}   & 0.265  & 6.816  \\
          &            & {CHOH}     & 0.609  & -1.824 \\
          &            & {OCHO}     & -0.125 & 2.820  \\
          &            & {OHCH$_3$} & -1.589 & -5.615 \\ \bottomrule
\end{tabular}
\caption{Comparison of adsorption energies predicted by the GNN to those computed with DFT, for the methanol pathway.}
\label{tab:dft_results}
\end{table}

\subsection{Analysis of Search Depth}
\label{sec:Analysis of Search Depth}

Beyond analyzing the final reward, it is also valuable to understand which of the \ourSystem search methods discover their best recommended catalysts with fewer computations. One aspect of of this search is the depth at which the maximum reward node is found. As shown in Table \ref{tab:search depth}, GPT-4 based searches find their maximum reward node in fewer nodes, on average than, GPT-3.5-turbo based searches, in each dataset. Then, \ourSystem-Expert shows superior performance in the OpenCatalyst and BioFuels datasets.

\begin{table*}
    \centering
    \caption{Comparison of the average search depth of best node in each of the search trees.}
    \begin{tabular}{ccccc}
        Method & Category & GPT-3.5-turbo & GPT-4 & Search depth reduction (\%)\\
        \hline
        \ourSystem-Expert & OpenCatalyst & 3.87 & 3.45 & 10.97 \\
        \ourSystem-Expert & BioFuels & 3.71 & 3.50 & 5.76\\
        \ourSystem-Expert & \ce{CO2}-Fuel & 4.30 & 3.30 & 23.25\\
        \hline
        \ourSystem-Planner & OpenCatalyst & 3.79 & 3.67 & 3.08\\
        \ourSystem-Planner & BioFuels & 3.5 & 3.21 & 8.16 \\
        \ourSystem-Planner & \ce{CO2}-Fuel & 4.25 & 3.55 & 16.47 \\
        \hline
    \end{tabular}
    \label{tab:search depth}
\end{table*}

\label{sec:Distribution of Reward Functions}

\begin{figure}
    \centering
    \includegraphics[width=0.95\linewidth]{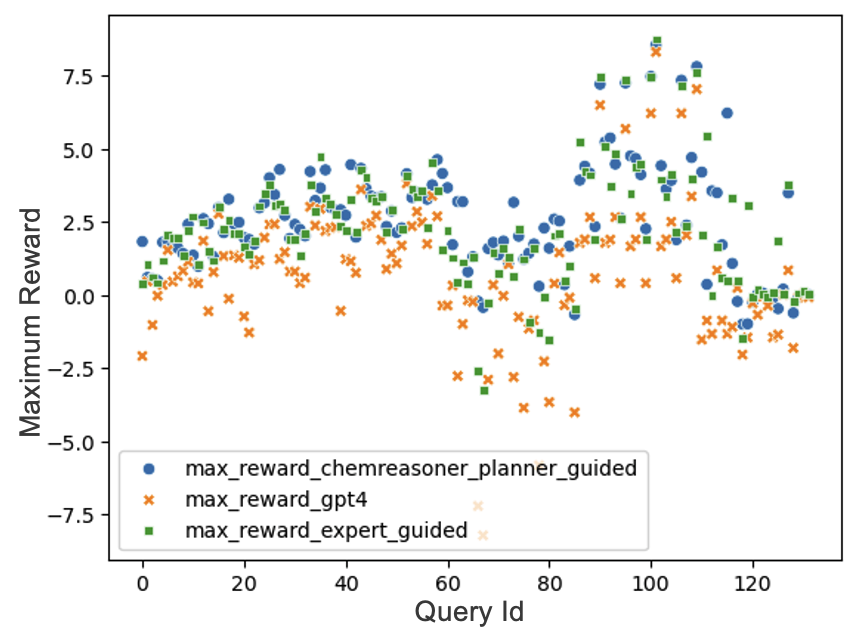}
    \caption{Distribution of simulation rewards for GPT-4 vs the two {\ourSystem} strategies.  {\ourSystem} consistently recommends better catalyst candidates across different simulations.}
    \label{fig:reward_scatter_plots}
\end{figure}

\subsection{Ablation on Actions Taken}

\begin{figure}
    \centering
    \includegraphics[width=1\linewidth]{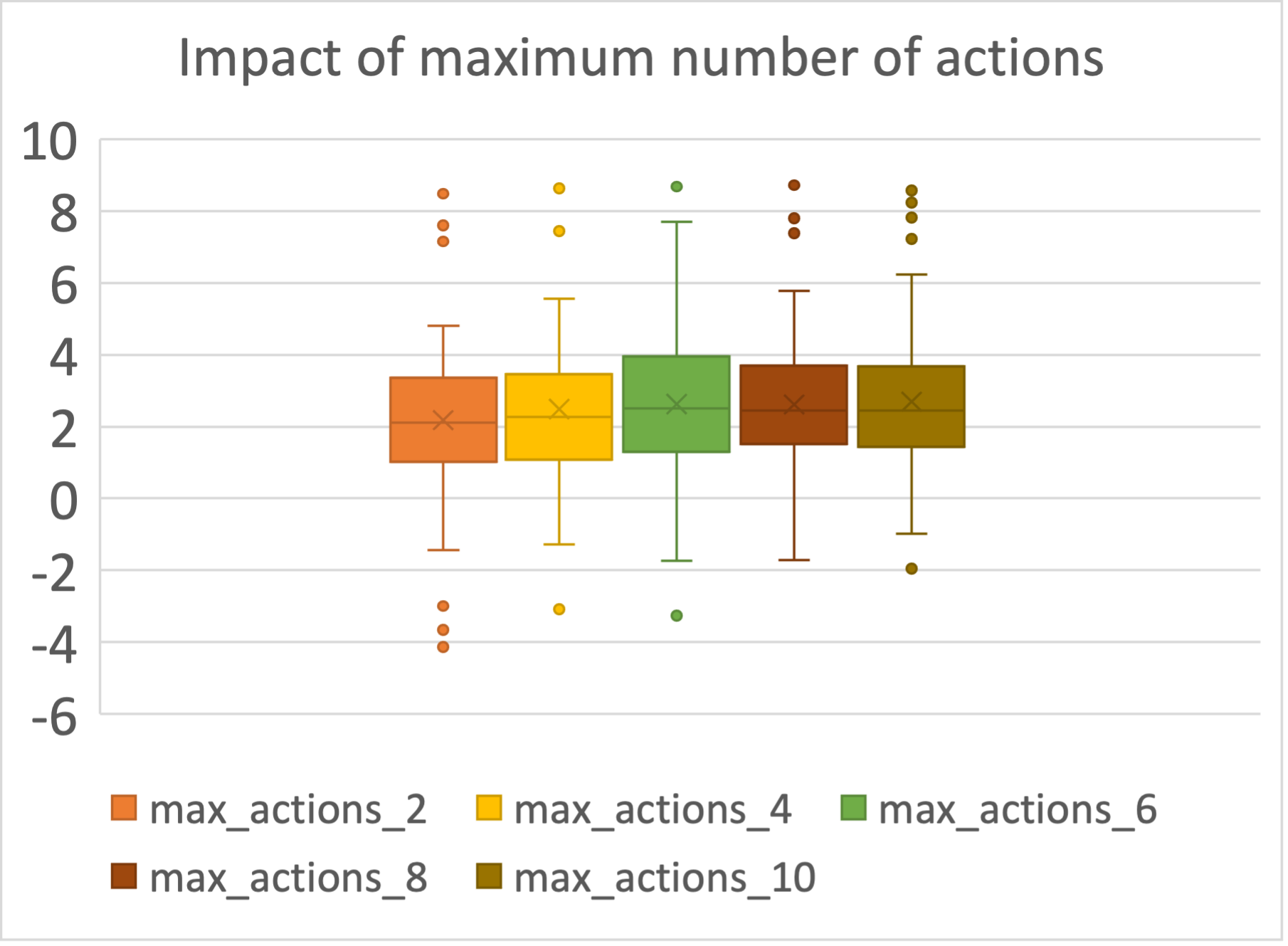}
    \caption{Impact of maximum number of actions on candidate search. Maximum actions if the most actions (e.g., ``exclude low selectivity'') that the LLM is allowed to consider at once when it is reasoning over catalysts.}
    \label{fig:num_actions}
\end{figure}

To further investigate \ourSystem's performance, we perform an ablation over the number of actions suggested from the planning prompt. While it may be unpredictable how many actions the LLM may propose in its plan, we can limit the number of actions, by randomly down-selecting actions, determining the ideal number of actions  for $\ourSystem$-Planner. In Figure \ref{fig:num_actions}, we show box-and-whisker plots of the best rewards for $\ourSystem$-Planner evaluated over our datasets for various maximum numbers of actions (10 was used in the main text). The final rewards are steadily increasing for max number of actions 2-6, while 8 and 10 are offer the highest rewards overall.

\subsection{Scaling of LLM and GNN inference}
\label{app:sec:Scaling of LLM and GNN inference}

Both GPT-4 and GPT-3.5-turbo were run using the Python OpenAI Azure interface. Our runtime was dominated by per-deployment rates limits, which are 40K tokens/minute and 240 requests/minute for GPT-4 and 300k tokens/minute and 1800 requests/minute for GPT-3.5-turbo. Prompts at each level in the heuristic search tree are executed asynchronously and in parallel.

Our implementation of the GNN reward function can perform an L-BFGS step in approximately 0.5s with a batch size of 40 adsorbate+catalyst configurations on a single GPU on a DGX2/V100 and on a single A100 40GB.

\section{Multi-modal models for Chemistry} \label{sec:mm_llm4chem}

Recent advances in NLP have produced strong results in the chemistry domain by training LLMs \citep{fabian2020molecular, chithrananda2020chemberta, vaucher2021inferring, schwaller2021mapping, megamolbartv2, tysinger2023can} on molecular sequence representations \cite{weininger1988smiles, weininger1989smiles, krenn2020self, cheng2023group}. Beyond this, multi-modal text-molecule models \cite{edwards2021text2mol, vall2021bioassayclr, zeng2022deep, xu2022protranslator, su2022molecular, seidl2023enhancing, xu2023protst, zhao2023adversarial, liu2023text} have been proposed to enable high-level control over molecular design \cite{zhang2023artificial}. This line of work largely focuses on cross-modal retrieval \cite{edwards2021text2mol, zeng2022deep}, translation \cite{edwards2022translation, liu2023molxpt, christofidellis2023unifying}, and editing \cite{liu2022multi}. Building on this trend, recent work has also emerged to enable this behavior via instruction-following \cite{zhao2023gimlet, fang2023mol, guo2023can} and dialogue-based systems for molecules \cite{liu2023chatgpt, cao2023instructmol, ye2023drugassist, zhao2024chemdfm}.
\section{Example of Automated Planner-driven Execution}
\label{sec:planner_trace}
In this example, we demonstrate the reasoning pathway followed by $\ourSystem$-Planner from the root node in a search tree to the best answer.

We provide the question and then present the internal reasoning generated by the LLM to determine the next search criteria.  We interleave each planning phase with the question and answers.

\textbf{Question}: Generate a list of top-5 {catalyst-label} that exhibit high activity for the CO2 to methanol conversion reaction. Provide scientific explanations and return a list of top 5 answers and their explanations as a list of pairs. Let's think step-by-step.

\subsection*{Level 0: Answer from User-specified Query}

1. Copper (Cu): Cu-based catalysts are the most widely used in the CO2 to methanol conversion process, mainly due to their high catalytic activity and selectivity towards methanol. The reaction mechanism involves the initial dissociation of CO2 on the Cu surface into CO and O, followed by the hydrogenation of CO to form methanol.

2. Zinc (Zn): Zinc is commonly used in conjunction with copper (Cu/Zn) for the CO2 to methanol conversion. Zinc does not directly participate in the catalysis but enhances the dispersion of copper particles, which increases the number of active sites and thereby the catalytic activity.

3. Palladium (Pd): Palladium can catalyze the CO2 to methanol conversion reaction via a formate pathway. The CO2 molecule adsorbs on the palladium surface and is reduced to formate, which subsequently gets hydrogenated to produce methanol.

4. Ruthenium (Ru): Ruthenium-based catalysts exhibit high activity for the CO2 to methanol conversion reaction. The CO2 molecule adsorbs on the Ru surface and undergoes dissociation into CO and O, which are then hydrogenated to form methanol.

5. Nickel (Ni): Nickel-based catalysts can efficiently catalyze the CO2 to methanol conversion via a CO pathway. The CO2 molecule initially dissociates into CO and O on the Ni surface, followed by the hydrogenation of CO to form methanol.

\subsection*{Level 1: Internal monologue of Planner}
To address the root question, we need to modify the search state to find metallic catalysts that are effective in the CO2 to methanol conversion reaction. Given the existing state, we need to add inclusion criteria that align with this requirement. 

Let's consider the following factors:

1) Catalyst Type: The root prompt asks for metallic catalysts, so we don't need to change the type of catalyst. However, to narrow down our search, we could specify a subclass of metallic catalysts that are known to be effective in this reaction. For example, transition metal catalysts such as copper, zinc, and palladium are known to be active in CO2 to methanol conversion.

2) Inclusion Criteria: Since we're looking for catalysts that exhibit high activity for the CO2 to methanol conversion reaction, we should include this as a criteria. We might also want to include catalysts that are stable under reaction conditions and have a high selectivity towards methanol.

3) Exclusion Criteria: Since we're looking for metallic catalysts, we can exclude non-metallic catalysts. Additionally, we might want to exclude catalysts that degrade quickly or are toxic.

4) Relationship to Candidate List: We're looking for catalysts that are similar to those in our candidate list. This means they should have similar properties and performance in the CO2 to methanol conversion reaction.

Here are my suggestions:

- Catalyst Type: ["transition metal catalysts"]
- Inclusion Criteria: ["high activity in CO2 to methanol conversion", "stability under reaction conditions", "high selectivity towards methanol"]
- Exclusion Criteria: ["non-metallic catalysts", "quickly degrading catalysts", "toxic catalysts"]
- Relationship to Candidate List: ["similar to"]

This way, we can find catalysts that are not only similar to those in our candidate list, but also meet the specific requirements of the root question.

\subsection*{Level 1: Answer from Planner-generated Query}

Based on the requirement of high activity for the CO2 to methanol conversion reaction, the following metallic catalysts are recommended:

1. **Rhodium (Rh)**: Rhodium-based catalysts have shown excellent performance in CO2 hydrogenation to methanol. The high activity of Rh is attributed to its ability to dissociate the C-O bond in CO2, which is a key step in the conversion process. Additionally, Rh has an appropriate binding energy for CO2, allowing for efficient adsorption and desorption of the reactant and product.

2. **Platinum (Pt)**: Platinum is known for its high catalytic activity in various reactions. In the case of CO2 to methanol conversion, Pt can activate the CO2 molecule by weakening the C-O bond, facilitating the subsequent hydrogenation steps. 

3. **Iridium (Ir)**: Iridium catalysts are effective in CO2 hydrogenation to methanol due to their high resistance to CO poisoning, a common issue in these reactions. Furthermore, Ir has a suitable binding energy for CO2, which enhances the reaction rate by promoting the adsorption and desorption processes.

4. **Gold (Au)**: Gold catalysts have recently been found to be effective in CO2 to methanol conversion. The high activity of Au is associated with its ability to activate CO2 by facilitating the breaking of the C-O bond. Also, Au has a high resistance to CO poisoning.

5. **Silver (Ag)**: Silver is another promising catalyst for CO2 to methanol conversion. The strong interaction between Ag and CO2 ensures efficient activation of the CO2 molecule, leading to high reaction rates. Moreover, Ag is known for its resistance to sintering, which maintains the catalyst's stability and longevity.

\subsection*{Level 2: Internal monologue of Planner}

Based on the current search state, answer and the root question, it is clear that we are looking for metallic catalysts that exhibit high activity for the CO2 to methanol conversion reaction. The inclusion criterion is commercial viability, and there are no exclusion criteria. The recommended catalysts should be similar to the ones provided in the candidate list.

From the current answer, we see that the recommended catalysts have properties such as high resistance to CO poisoning, efficient activation of CO2, ability to dissociate the C-O bond in CO2, suitable binding energy for CO2, high resistance to sintering, etc. 

Based on these, here are some possible actions that we can take:

- **Catalyst Type:** We can look for other types of metallic catalysts that have not been included in the current list. For instance, we can search for "transition metal catalysts" or "noble metal catalysts", since these categories often exhibit high catalytic activities and have similar properties to the catalysts in the current list.

- **Inclusion Criteria:** We can add new inclusion criteria based on the properties of the recommended catalysts in the current answer. For example, we can include "high resistance to CO poisoning" or "ability to dissociate the C-O bond in CO2".

- **Exclusion Criteria:** Since all the recommended catalysts in the current answer seem to have desirable properties, it may not be necessary to add any exclusion criteria at this point. However, if we want to narrow down our search, we could consider excluding catalysts that have "low resistance to sintering" or "poor binding energy for CO2".

- **Relationship to Candidate List:** The relationship to the candidate list can remain as "similar to", as we are looking for catalysts that have similar properties to the ones in the current list.

"catalyst-type": ["transition metal catalysts", "noble metal catalysts"],
 "inclusion-criteria": ["high resistance to CO poisoning", "ability to dissociate the C-O bond in CO2"],
 "exclusion-criteria": ["low resistance to sintering", "poor binding energy for CO2"],
 "relationship-to-candidate-list": ["similar to"]

\subsection*{Level 2: Answer from Planner-generated Query}

Sure, here is the list of top-5 metallic catalysts that exhibit high activity for the CO2 to methanol conversion reaction with properties of commercial viability and high resistance to CO poisoning:

1. Copper (Cu): Copper-based catalysts, particularly when combined with zinc oxide and alumina (Cu/ZnO/Al2O3), are the most widely used in the industrial production of methanol from CO2. Copper has a high catalytic activity and selectivity towards methanol production. The resistance to CO poisoning is improved by the presence of ZnO, which can absorb CO2 and present it to the copper sites in a more reactive form.

2. Palladium (Pd): Palladium catalysts, particularly when supported on carbon (Pd/C), show high activity for CO2 reduction to methanol. Pd has the ability to dissociate CO2 into CO and O, facilitating the subsequent hydrogenation of CO to methanol. The carbon support provides resistance to CO poisoning by facilitating the desorption of CO from the Pd surface.

3. Ruthenium (Ru): Ruthenium catalysts, especially when supported on carbon (Ru/C), show good activity and selectivity for CO2 reduction to methanol. Ru has a high CO dissociation energy, which helps resist CO poisoning. The carbon support helps disperse the Ru particles to maximize the catalytic surface area.

4. Nickel (Ni): Nickel catalysts, particularly when supported on alumina (Ni/Al2O3), have been shown to catalyze the CO2 to methanol reaction. Nickel has a lower CO binding energy than other metals, which helps resist CO poisoning. The alumina support helps disperse the Ni particles and stabilize the catalyst.

5. Cobalt (Co): Cobalt catalysts, particularly when supported on ceria (Co/CeO2), can catalyze the CO2 to methanol reaction. Cobalt has a relatively low CO binding energy, which helps resist CO poisoning. The ceria support can enhance the CO2 adsorption and activation, improving the catalytic activity.

Comparing to the initial list, these metals are generally cheaper and more abundant, making them more commercially viable. They also show a good resistance to CO poisoning due to their lower CO binding energies and/or the use of specific supports.

\subsection*{Level 3: Internal monologue of Planner}
 Based on the \$searchState, \$rootPrompt, \$currentQuestion, and \$currentAnswer, here are some possible action suggestions:

1) "catalyst-type": The current catalyst type is "metallic catalysts". The \$currentAnswer provides a list of metallic catalysts that are commercially viable and have high resistance to CO poisoning. Based on this, we could further specify the type of metallic catalysts to explore. For example, we could look into "transition metal catalysts" or "noble metal catalysts". These categories are subclasses of metallic catalysts and could potentially offer new candidates that have not been considered yet. 

2) "inclusion-criteria": The current inclusion criteria are "commercial viability" and "high resistance to CO poisoning". The \$currentAnswer explains that these properties are achieved through the choice of metal and the use of specific supports. This suggests that we could add criteria related to these factors. For instance, we could add "use of specific supports to enhance catalytic activity and resistance to CO poisoning" as a new inclusion criteria.

3) "exclusion-criteria": The current exclusion criteria is an empty list. Based on the \$currentAnswer, all the suggested catalysts are not only commercially viable and resistant to CO poisoning, but also have good catalytic activity for the CO2 to methanol conversion reaction. Therefore, we could add "low catalytic activity for the CO2 to methanol conversion reaction" as an exclusion criterion.

4) "relationship-to-candidate-list": The current relationship to the candidate list is "similar to". Since the \$currentAnswer provides a new list of catalysts that are different from the initial list, we could change the relationship to "different from".

"catalyst-type": ["transition metal catalysts", "noble metal catalysts"],
 "inclusion-criteria": ["use of specific supports to enhance catalytic activity and resistance to CO poisoning"],
 "exclusion-criteria": ["low catalytic activity for the CO2 to methanol conversion reaction"],
 "relationship-to-candidate-list": ["different from"]

 These changes would allow us to explore new categories of metallic catalysts while focusing on the important properties and excluding those with low activity for the CO2 to methanol conversion reaction.

\subsection*{Level 3: Answer from Planner-generated Query}
Let's begin by analyzing the original list of catalysts:

1. Copper (Cu): Commercially viable and shows high resistance to CO poisoning. It has been extensively used in industry for the conversion of CO2 to methanol due to its high catalytic activity. Cu-based catalysts, especially when combined with zinc oxide and alumina, exhibit remarkable performance.

2. Palladium (Pd): While Pd-based catalysts show high catalytic activity for many reactions, they are known to be susceptible to CO poisoning, which could limit their effectiveness in the CO2 to methanol conversion.

3. Ruthenium (Ru): Ru-based catalysts exhibit high catalytic activity and resistance to CO poisoning. However, their high cost and limited availability may hinder their commercial viability.

4. Nickel (Ni): Ni-based catalysts are commercially viable due to their relatively low cost. However, they exhibit moderate activity for the CO2 to methanol conversion and could be susceptible to CO poisoning.

5. Cobalt (Co): Co-based catalysts are less active for the CO2 to methanol conversion compared to the other catalysts listed. Also, they are susceptible to CO poisoning.

Based on the above analysis, let's create a new list of metallic catalysts that meet the specified requirements:

1. Rhodium (Rh): Rh-based catalysts show high catalytic activity for the CO2 to methanol conversion and have high resistance to CO poisoning. However, their high cost may limit their commercial viability.

2. Silver (Ag): Ag-based catalysts exhibit high catalytic activity for the CO2 to methanol conversion, and are resistant to CO poisoning. They are also commercially viable due to their relative abundance and lower cost compared to noble metals.

3. Iron (Fe): Fe-based catalysts are commercially viable due to their low cost and abundance. They show moderate to high activity for the CO2 to methanol conversion and have good resistance to CO poisoning.

4. Platinum (Pt): Pt-based catalysts have high catalytic activity and resistance to CO poisoning. However, their high cost may limit their commercial viability.

5. Gold (Au): Au-based catalysts show high catalytic activity for the CO2 to methanol conversion and have high resistance to CO poisoning. However, their high cost may limit their commercial viability.

This list includes catalysts that exhibit high catalytic activity for the CO2 to methanol conversion and high resistance to CO poisoning. However, the commercial viability of some of these catalysts may be limited due to their high cost.

\subsection*{Level 4: Internal monologue of Planner}

 From the \$currentAnswer, it seems that we have already achieved the intent of the \$rootPrompt. We have a list of top-5 metallic catalysts that exhibit high activity for the CO2 to methanol conversion reaction, and have given scientific explanations for each of the metallic catalysts. The list has also been compared to the initial list of catalysts.

However, if we wanted to refine the search further, we could consider the following actions:

1) "Change the type of catalyst to search for": We could limit our search to only noble metallic catalysts. Noble metals like Rhodium, Platinum, and Gold have already been identified as having high catalytic activity and resistance to CO poisoning. However, their commercial viability may be limited due to their high cost. Including less costly noble metals like Silver might provide a good balance between catalytic performance and commercial viability.

2) "Add a new inclusion criteria": We could add "relative abundance" as a new inclusion criteria. This would ensure that we only consider catalysts that are not only commercially viable and resistant to CO poisoning, but also readily available.

3) "Add a new exclusion criteria": We could add "high cost" as a new exclusion criteria. This would help us exclude catalysts that are not commercially viable, despite their high catalytic activity and resistance to CO poisoning.

4) "Change the relationship to the candidate list": Given that we already have a list of suitable catalysts, we could modify the relationship to be "different from". This would help us identify new and potentially better catalysts.

Here's how we could represent these suggested actions:

 "catalyst-type": ["noble metallic catalysts"],
 "inclusion-criteria": ["relative abundance"],
 "exclusion-criteria": ["high cost"],
 "relationship-to-candidate-list": ["different from"]

These suggestions would help us refine our search and potentially identify new metallic catalysts that are commercially viable, resistant to CO poisoning, readily available, and not excessively costly.

\subsection*{Level 4: Answer from Planner-generated Query}

The list provided contains both noble (Rh, Ag, Pt, Au) and non-noble (Fe) metallic catalysts. For the CO2 to methanol conversion, a noble metallic catalyst is desirable for its high resistance to CO poisoning which is important as CO is an intermediate in this reaction. Iron, however, is not a noble metal and does not have this property. 

1. Palladium (Pd): Palladium is a noble metal that can be used for CO2 to methanol conversion. It has high resistance to CO poisoning. It is also commercially viable since it is widely used in the industry for hydrogenation reactions. Moreover, palladium can dissociate CO2 into CO and O, which is beneficial for methanol synthesis.

2. Ruthenium (Ru): Ruthenium is another noble metal that can be used for this conversion. It shows high resistance to CO poisoning and is commercially viable. Ruthenium can adsorb CO2 and hydrogen simultaneously, facilitating the conversion to methanol.

3. Rhodium (Rh): Rhodium is a noble metal with high catalytic activity for the CO2 to methanol conversion. It is resistant to CO poisoning and is commercially viable. It has an excellent ability to dissociate CO2 and hydrogen, facilitating the conversion to methanol.

4. Platinum (Pt): Platinum is a noble metal that is highly resistant to CO poisoning. It is commercially viable and is often used in catalysis due to its high reactivity. Platinum can dissociate CO2 into CO and O, which is beneficial for methanol synthesis.

5. Gold (Au): Gold is a noble metal that is resistant to CO poisoning. It is commercially viable and is known to catalyze the CO2 to methanol conversion, especially when used as nanoparticles. Gold can adsorb and activate CO2, facilitating its conversion to methanol.

Therefore, the updated list of noble metallic catalysts suitable for the CO2 to methanol conversion, considering the given criteria, is ['Palladium (Pd)', 'Ruthenium (Ru)', 'Rhodium (Rh)', 'Platinum (Pt)', 'Gold (Au)'].

\subsection*{Level 5: Internal monologue of Planner}
Looking at the \$rootPrompt, the task is to generate a list of top-5 metallic catalysts. The \$currentAnswer provides a list of noble metallic catalysts. However, the \$rootPrompt does not restrict the type of metallic catalysts to noble ones. Therefore, one possible action is to expand the search to include non-noble metallic catalysts.

Furthermore, the \$currentAnswer includes scientific explanations for each catalyst. These explanations can be used to refine the inclusion and exclusion criteria. For example, the ability to dissociate CO2 into CO and O is a desirable property mentioned multiple times in the explanations. Therefore, this can be added as a new inclusion criterion. On the other hand, low catalytic activity for the CO2 to methanol conversion reaction is the current exclusion criterion. We can add another exclusion criterion based on the explanations provided. For example, we can exclude catalysts that cannot adsorb and activate CO2, as this property is also mentioned as being beneficial for the conversion to methanol.

The relationship to the candidate list can be changed to "different from" to find new metallic catalysts that are not in the current list.

Here are the suggestions based on the above analysis:

 "catalyst-type": ["non-noble metallic catalysts"],
 "inclusion-criteria": ["ability to dissociate CO2 into CO and O", "ability to adsorb and activate CO2"],
 "exclusion-criteria": ["cannot adsorb and activate CO2"],
 "relationship-to-candidate-list": ["different from"]

These changes will help in finding new metallic catalysts that are not noble metals but still have high catalytic activity for the CO2 to methanol conversion reaction. This is in line with the intent of the \$rootPrompt.

\subsection*{Level 5: Answer from Planner-generated Query}

The CO2 to methanol conversion reaction is a crucial process in catalysis and has been extensively studied. Here are my top-5 recommendations for noble metallic catalysts that exhibit high activity for this reaction, along with the scientific explanations:

1. **Palladium-Gold (Pd-Au) Alloy**: This alloy catalyst has shown excellent catalytic performance for CO2 reduction to methanol. The Pd-Au alloy, especially when supported on titanium dioxide (TiO2), provides a synergistic effect where Pd facilitates CO2 activation and dissociation, while Au helps in reducing the CO intermediate to methanol. The alloy structure also enhances the resistance to CO poisoning.

2. **Platinum-Ruthenium (Pt-Ru) Alloy**: Pt-Ru alloy, when supported on carbon (C), exhibits high activity and selectivity for CO2 to methanol conversion. The bifunctional mechanism involves CO2 activation and dissociation on Ru sites and subsequent hydrogenation to methanol on Pt sites. 

3. **Ruthenium-Gold (Ru-Au) Alloy**: The Ru-Au alloy is another excellent choice for this reaction. Ru helps in dissociating CO2 into CO and O, while Au aids in the reduction of CO to methanol. The alloy structure also enhances the resistance to CO poisoning.

4. **Rhodium-Palladium (Rh-Pd) Alloy**: The Rh-Pd alloy, when supported on alumina (Al2O3), exhibits high activity for CO2 to methanol conversion. Rh facilitates CO2 dissociation, while Pd helps in reducing the CO intermediate to methanol. The alloy structure enhances the resistance to CO poisoning.

5. **Platinum-Gold (Pt-Au) Alloy**: The Pt-Au alloy, when supported on carbon (C), shows high activity and selectivity for the CO2 to methanol conversion. The bifunctional mechanism involves CO2 activation and dissociation on Au sites and subsequent hydrogenation to methanol on Pt sites.
\end{document}